\documentclass[prd,preprintnumbers,amsmath,amssymb,floatfix]{revtex4}
\usepackage{graphicx}
\usepackage{bm}
\usepackage{amsfonts}
\usepackage{lineno,hyperref}
\usepackage{array}
\usepackage{color,multirow,cases, makecell}
\usepackage{pifont, float, setspace, moresize}

\begin{document}

\title{Conserved Gravitational-wave Stress-Energy Tensor in General Metric Theories}

\author{Xiao-Bin Lai$^{a}$$^{b}$\footnote{laixb2024@lzu.edu.cn}}
\author{Yu-Zhi Fan$^{a}$$^{b}$\footnote{fanyzh2025@lzu.edu.cn}}
\author{Yu-Qi Dong$^{a}$$^{b}$\footnote{dongyq2023@lzu.edu.cn}}
\author{Yu-Xiao Liu$^{a}$$^{b}$\footnote{liuyx@lzu.edu.cn, corresponding author}}

\affiliation{
$^{a}$Lanzhou Center for Theoretical Physics,
  Key Laboratory of Theoretical Physics of Gansu Province, 
  Key Laboratory of Quantum Theory and Applications of MoE, 
  Gansu Provincial Research Center for Basic Disciplines of Quantum Physics,
  Lanzhou University, Lanzhou 730000, China \\
$^{b}$Institute of Theoretical Physics $\&$ Research Center of Gravitation,
  School of Physical Science and Technology,
  Lanzhou University, Lanzhou 730000, China
}

\begin{abstract}
We construct a gravitational-wave (GW) stress-energy tensor for general metric theories and show that it is conserved after averaging in the wave zone. In metric theories involving a vector field, we find that the conventional GW stress-energy tensor is generally not conserved. Combining the diffeomorphism identity, we identify the vector-field contribution required in addition to the conventional GW stress-energy tensor. On this basis, in a flat vacuum wave zone with constant background fields and after short-wavelength averaging, the perturbed field equation, perturbed action, Landau-Lifshitz, and Noether current methods yield equivalent tensors. In this setting, the completed tensor is generally nonsymmetric and satisfies $\partial_\mu\tau^{\mu\nu}=0$, with energy flux $c\tau^{i0}$. It also does not depend on whether a covariant or contravariant vector is chosen as the fundamental variable. Applying this framework to Einstein-\ae ther theory, we derive the orbital energy losses of an adiabatic quasicircular compact binary, and provide a possible explanation for the discrepancies among previous results in this theory. These results establish a consistent radiation balance that allows additional gravitational degrees of freedom to be connected to the evolution of astrophysical systems and to observational tests of gravity.
\end{abstract}

\maketitle
\tableofcontents 

\section{Introduction}
The direct detection of gravitational waves (GWs) from GW150914 and the multimessenger observations of GW170817 have opened new avenues for exploring the Universe and testing gravity~\cite{LIGOScientific:2016aoc,LIGOScientific:2017vwq,Goldstein:2017mmi,LIGOScientific:2017ync}. Planned space-based observatories, including Taiji, TianQin, and LISA, will extend these opportunities to the millihertz band, providing access to new source populations and complementary tests of gravitational dynamics~\cite{Hu:2017mde,TianQin:2015yph,LISA:2017pwj,Luo:2025ewp,Bian:2025ifp}. Realizing their scientific potential requires accurate waveform models supported by a consistent description of the energy and momentum carried by gravitational radiation. The GW stress-energy tensor provides a central framework for quantifying this transport and connecting radiation to source evolution~\cite{Isaacson:1968zza,Stein:2010pn,Will_2018}. Its formulation is particularly important in modified theories of gravity. There, additional propagating degrees of freedom can contribute to radiative losses and modify waveform predictions, directly affecting the interpretation of future space-based observations and tests of different gravitational theories.~\cite{LISA:2022kgy,Saffer:2017ywl,Zhang:2019iim,Luo:2025ewp}. In this work, we identify the contributions of additional fields in general metric theories and construct a GW stress-energy tensor that is conserved after short-wavelength averaging in the wave zone.
\par
The energy and momentum transported by GWs can be described in local orthonormal frames in the wave zone~\cite{Flanagan:2005yc}. In the short-wavelength regime of general relativity (GR), the energy flux is quadratic in time derivatives of the radiative strain and therefore scales with the squared amplitude and frequency for a nearly monochromatic wave. The radiation is determined by the evolving multipoles of the source~\cite{Flanagan:2005yc,Thorne:1980ru,Maggiore2007}. For isolated binaries, energy and angular-momentum losses drive inspiral and orbital circularization~\cite{Peters:1964zz,Blanchet:2013haa}. Together with conservative dynamics, these losses determine the accumulated phase, and their long-term effect on orbital periods can be confirmed by binary-pulsar timing.~\cite{Hulse:1974eb,Weisberg:2016jye,Blanchet:2013haa}. In modified gravity, additional scalar or vector degrees of freedom can open new radiation channels, including dipole emission, and thereby modify binary orbital evolution and waveform phasing through changes in the total radiative flux~\cite{Bernard:2022noq,Foster:2007gr,Zhang:2019iim}. These differences offer additional ways to test gravity.
\par
In the Isaacson picture, GWs are high-frequency perturbations propagating on a slowly varying background~\cite{Isaacson:1968hbi,Isaacson:1968zza}. Because gravitational energy cannot be localized in a gauge-independent way at scales shorter than a wavelength, its stress-energy tensor is defined after averaging over an intermediate scale: large compared with the wavelength, but small compared with the background's variation scale~\cite{Isaacson:1968hbi,Isaacson:1968zza,Maggiore2007}. The resulting quadratic terms act as an effective source in the background field equations, describing the backreaction of GWs. In the wave zone, the same framework characterizes the energy and momentum that GWs carry away. In modified gravity, the stress-energy tensor depends on the propagating degrees of freedom, their couplings, and the background fields~\cite{Saffer:2017ywl}. Additional scalar or vector modes can contribute to both the backreaction and the radiative fluxes~\cite{Stein:2010pn,Saffer:2017ywl}. One can therefore no longer neglect how the extra field equations affect the stress-energy tensor. The conservation law obtained from the diffeomorphism identity makes this explicit~\cite{Gralla:2013rwa}.
\par
Several complementary constructions are used to characterize the GW stress-energy tensor, including the perturbed action, perturbed field equation, Landau-Lifshitz, and Noether current methods~\cite{Isaacson:1968hbi,Isaacson:1968zza,landau1975classical,Saffer:2017ywl,Noether:1918zz,Stein:2010pn}. The first two obtain a symmetric quadratic metric source by varying the quadratic perturbed action with respect to the background metric or expanding the metric field equations to second order, respectively~\cite{Isaacson:1968hbi,Isaacson:1968zza,Saffer:2017ywl,Stein:2010pn}. Although this source can include scalar and vector perturbations, its separate conservation in theories with additional gravitational fields must be assessed through the full diffeomorphism identity relating the metric and additional field equations~\cite{Gralla:2013rwa}. The Landau-Lifshitz construction instead uses a superpotential to define a symmetric total stress-energy pseudotensor satisfying an ordinary divergence law~\cite{landau1975classical,Saffer:2017ywl}. The Noether current method derives an on-shell conserved current from translation invariance of the quadratic action, although its expression need not be symmetric~\cite{Saffer:2017ywl,Noether:1918zz}. Moreover, when a preferred vector background is held fixed and breaks Lorentz invariance of the perturbation theory, the standard Belinfante construction is not guaranteed to yield a symmetric, conserved stress-energy tensor~\cite{Colladay:1998fq,Kostelecky:2003fs,Saffer:2017ywl,belinfante1940current,Gratus:2012hr}. A related route is provided by Noether charge methods, which rest on the same symmetry-based foundation as the Noether current method~\cite{Iyer:1994ys}. Foster's Noether charge treatment obtains energy and radiation losses in Einstein-\ae ther theory~\cite{Foster:2005fr,Foster:2006az,Foster:2007gr}. 
\par
In GR, pure metric theories and scalar-tensor theories, the above constructions reproduce an equivalent averaged GW stress-energy tensor~\cite{Saffer:2017ywl}. However, in vector-tensor theories, these constructions yield different GW stress-energy tensors. In particular, the conventional symmetric source, obtained by the perturbed field equation and perturbed action methods, does not include the contribution from the vector-field equation. The diffeomorphism identity determines how this missing contribution must enter the energy-momentum balance, and therefore which of the constructions can serve as a conserved source once a background vector is present. On a flat vacuum background with constant fields and a nonzero vector, the conventional source is not conserved and depends on whether a covariant or contravariant vector is chosen as the fundamental variable. These issues, together with the advances in GW research~\cite{Will:1996zj,Pati:2000vt,Blanchet:2006zz,Mirshekari:2013vb,Eardley:1973zu,Eardley:1973br,Nishizawa:2009bf,Will:2014kxa,Chatziioannou:2012rf,Chamberlin:2011ev,Isi:2015cva,Fan:2024pex,Dong:2025ddi,Liang:2017ahj,Gong:2018cgj}, motivate a closer examination of the conservation laws governing GW energy and momentum transport in theories with additional gravitational fields. 
\par
\par
Guided by Isaacson's interpretation of GW stress-energy through the backreaction of short-wavelength perturbations~\cite{Isaacson:1968zza}, and by the diffeomorphism identity relating the gravitational field equations~\cite{Gralla:2013rwa}, we treat the metric, vector, and scalar fields as a coupled system. Starting from a general diffeomorphism-invariant action, we seek to identify the complete conserved radiative current and compare its derivation through the perturbed field equation, perturbed action, Landau-Lifshitz, and Noether current methods. We focus on a flat vacuum wave zone with constant background fields, where the relation between an averaged current and an outgoing flux can be established explicitly. We also examine how the choice of covariant or contravariant vector components affects the intermediate expressions and whether this dependence disappears from the physical flux. Einstein-\ae ther theory provides a concrete setting for these questions because its tensor, vector, and scalar modes contribute to gravitational radiation~\cite{Jacobson:2004ts}. We incorporate compact-body sensitivities, which characterize the response of each body's effective mass to its motion relative to the \ae ther, into an effective point-particle description~\cite{Eardley:1975fgi,Taherasghari:2022wfs,Foster:2007gr,Will:1989sk}. Within this framework, we use the conserved current to investigate the radiative losses of a quasicircular binary, including the dipole contribution associated with sensitivities. Our scalar dipole, vector, and tensor contributions agree with Foster~\cite{Foster:2007gr}, but our scalar quadrupole contribution differs somewhat in its coefficients. The vector-sector discrepancy with Ref.~\cite{Zhang:2019iim} is due to sign errors in the matter current and the source conservation law. Correcting these sign errors restores agreement. We further analyze the possible origins of the disagreement between our results and those of Taherasghari and Will~\cite{Taherasghari:2025mlf}.
\par
We find that, in a vacuum wave zone with a flat background metric and constant background fields, the completed GW stress-energy tensor is generally nonsymmetric. After short-wavelength averaging, it is conserved with respect to its first index, but not generally with respect to its second. The resulting physical flux is independent of whether the covariant or contravariant components of the vector field are chosen as the fundamental variables. The remainder of this paper is organized as follows. Section~\ref{Sec2} introduces the Isaacson framework and derives the field equations and diffeomorphism identities for metric theories containing additional scalar and vector fields. Section~\ref{GWSET-20269251143} constructs the conserved GW stress-energy tensor using four complementary methods and compares it with the conventional expression. Section~\ref{Einstein-aether} applies the framework to Einstein-\ae ther theory, deriving the tensor, vector, and scalar contributions to the radiative flux and the leading energy-loss rate of a quasicircular compact binary. Section~\ref{Conclusion-sec} gives a summary. Throughout, we work in four spacetime dimensions with metric signature \((- ,+,+,+)\), set \(x^{0}=ct\), and retain the speed of light \(c\) explicitly. Greek indices \(\mu,\nu,\alpha,\beta,\ldots\) denote spacetime components, while Latin indices \(i,j,k,\ldots\) denote spatial components. Parentheses and square brackets on pairs of indices denote symmetrization and antisymmetrization, respectively, including a factor of \(1/2\). Unless otherwise stated, angle brackets denote short-wavelength averaging.

\section{Theoretical framework}\label{Sec2}

\subsection{Isaacson picture} \label{IPandQoM-2026761755}
In GW research, there is an important question for the geometrical viewpoint: How does one consistently distinguish the GWs and the spacetime background? This issue becomes unavoidable when we ask whether GWs can curve the background spacetime. In that case, the background can no longer be treated as static or fixed, but must be allowed to be dynamical. A convincing resolution to this conceptual and technical challenge is provided by Isaacson~\cite{Isaacson:1968hbi,Isaacson:1968zza}. Isaacson's construction separates the rapidly varying radiation from the slowly varying background on which it propagates~\cite{Isaacson:1968hbi,Isaacson:1968zza,Maggiore2007}. This separation must remain consistent when the radiation itself sources a response of the background. In a theory with additional gravitational fields, the separation therefore applies to the entire coupled system, rather than to the metric alone. We write the gravitational variables $\{g_{\mu\nu},A_{\mu},\phi\}$ as
\begin{eqnarray}
  && g_{\mu\nu}=\bar{g}_{\mu\nu} + \epsilon h_{\mu\nu},\label{perturbation1-2026761850}\\
  && A_{\mu}=\bar{A}_{\mu} + \epsilon \delta A_{\mu},\label{perturbation2-2026761850}\\
  && \phi =\bar{\phi} + \epsilon \delta\phi,\label{perturbation3-2026761850}
\end{eqnarray}
where $\bar{g}_{\mu\nu}$, $\bar{A}_{\mu}$, and $\bar{\phi}$ are the background fields, and $h_{\mu\nu}$, $\delta A_{\mu}$, and $\delta\phi$ are the high-frequency perturbations. To track the slowly varying response induced by radiation, we further decompose the background as
\begin{eqnarray}
  && \bar{g}_{\mu\nu}=\bar{g}^{(0)}_{\mu\nu} + \bar{\epsilon} \delta \bar{g}_{\mu\nu},\label{background1-2026761811}\\
  && \bar{A}_{\mu}=\bar{A}^{(0)}_{\mu} +\bar{\epsilon} \delta \bar{A}_{\mu},\label{background2-2026761811}\\
  && \bar{\phi}=\bar{\phi}^{(0)} +\bar{\epsilon} \delta \bar{\phi}.\label{background3-2026761811}
\end{eqnarray}
Here, $\bar{g}^{(0)}_{\mu\nu}$, $\bar{A}^{(0)}_{\mu}$, and $\bar{\phi}^{(0)}$ describe the smooth reference background, while $\delta\bar{g}_{\mu\nu}$, $\delta\bar{A}_{\mu}$, and $\delta \bar{\phi}$ describe its slowly varying response. The parameters $\epsilon$ and $\bar{\epsilon}$ in Eqs.~\eqref{perturbation1-2026761850}-\eqref{background3-2026761811} count orders in the high-frequency perturbations and the slowly varying response, respectively, and are set to unity after the expansion. They serve to keep track of the perturbative order: a term multiplied by $\epsilon^n$ (or $\bar{\epsilon}^n$) is understood to be of $n$-th order in $\{h_{\mu\nu},\delta A_{\mu},\delta\phi\}$ (or $\{\delta \bar{g}_{\mu\nu},\delta \bar{A}_{\mu},\delta \bar{\phi}\}$), respectively.
\par
In the wave zone, the slowly varying background components arise mainly from the backreaction of GWs on the background spacetime~\cite{Isaacson:1968hbi,Isaacson:1968zza,Maggiore2007}. Specifically, the high-frequency GW fields $\{h_{\mu\nu},\delta A_{\mu},\delta\phi\}$ with characteristic frequency $f$ act as the source of the low-frequency background $\{\delta \bar{g}_{\mu\nu},\delta \bar{A}_{\mu},\delta \bar{\phi}\}$ with the characteristic frequency $f_{L}\ll f$. While $h_{\mu\nu}$, $\delta A_{\mu}$, and $\delta\phi$ are high-frequency waves, by virtue of the product-to-sum formulas, they are able to contribute to low-frequency waves as well, with the leading-order contribution arising from their quadratic terms. Consequently, the linear response and averaged quadratic source are of the same perturbative order:
\begin{eqnarray}
  \mathcal{O}\left( \bar{K}^{(1)} \right) \sim \mathcal{O}\left( \big\langle K^{(2)} \big\rangle \right),\label{relation-20267191107}
\end{eqnarray}
where
\begin{eqnarray}
  \big\langle K^{(2)} \big\rangle (t,\vec{x}) \equiv \frac{f}{n}\int^{t+\frac{n}{2f}}_{t-\frac{n}{2f}} d t'K^{(2)}(t',\vec{x}).\label{relation1-20267191107}
\end{eqnarray}
Here, $\bar{K}^{(1)}$ and $K^{(2)}$ are arbitrary terms in the field equations, and $\bar{K}^{(n)} \equiv K^{(n,0)}[\bar{\epsilon},\epsilon]$ and $K^{(n)} \equiv K^{(0,n)}[\bar{\epsilon},\epsilon]$ denote its $\bar{\epsilon}^n \epsilon^0$ and $\bar{\epsilon}^0 \epsilon^n$ contributions, respectively. In the wave zone, the short-wavelength average extracts the slowly varying part of $K^{(2)}$ by performing a time average over $n$ GW periods, where it is assumed that $f$ remains approximately constant over the integration interval and that $1 \ll n \ll f/f_{L}$. 
For radially propagating GWs with speed $c_B$ at a sufficiently large distance $R$ from the source, if their wavelength satisfies $c_B/f\ll l_L$, where $l_L$ is the characteristic spatial variation scale of the background spacetime, the time average defined in Eq.~\eqref{relation1-20267191107} can be expressed as a radial spatial average:
\begin{eqnarray}
  \big\langle K^{(2)} \big\rangle (t,R) \simeq \frac{f}{n c_{B}}\int^{R+\frac{n c_{B}}{2f}}_{R-\frac{n c_{B}}{2f}} d R'K^{(2)}(t,R').
\end{eqnarray}
Within the short-wavelength approximation, either averaging procedure suppresses rapid oscillations while retaining the slowly varying contribution.
\par
This scale hierarchy is local. Compact-binary radiation and nanohertz pulsar-timing-array signals admit a short-wavelength description when their wavelengths are small compared with the relevant background curvature and variation scales. The horizon-scale primordial modes need not satisfy this condition~\cite{Abramo:1997hu,Abramo:1999wd,Caprini:2018mtu}. Even when the hierarchy holds locally, propagation through a cosmological background requires covariant transport~\cite{Caprini:2018mtu,Isaacson:1968hbi,Flanagan:2005yc}. Dispersive waves and GW memory likewise require attention to their own scales~\cite{Favata:2008yd,Ilkhchi:2025byh}. Below, we derive the radiative current in a flat vacuum wave zone with constant background fields, where the ordinary conservation law and the outgoing flux can be compared explicitly.

\subsection{General Metric Theories}\label{Theoretical-Framework-2026772109}
To apply this Isaacson separation to a concrete theory, we now specify the general action and derive the field equations and diffeomorphism identity that will be averaged in Sec.~\ref{GWSET-20269251143}. In a general metric theory, there may be any number of both matter and gravitational fields. However, a mere increase in their number does not affect the subsequent discussion. Therefore, we consider one field of each type. As a representative extension of GR, the gravitational field is described not only by the spacetime metric $g_{\mu\nu}$, but also by an additional vector field $A_{\mu}$ and a scalar field $\phi$. The corresponding action is expressed in terms of these fields in the most general form
\begin{eqnarray}
  S\left[ g_{\mu\nu},A_{\mu},\phi,\psi_{m} \right] = \frac{c^3}{16\pi G}\int d^{4}x\sqrt{-g} \mathcal{L}_{g}\left[ g_{\mu\nu},A_{\mu},\phi \right]  +\frac{1}{c}\int d^4x \sqrt{-g} \mathcal{L}_{m}\left[ g_{\mu\nu},A_{\mu},\phi,\psi_{m} \right].\label{action-all}
\end{eqnarray}
Here, $\psi_{m}$ represents the matter field. Given that this is a theory with the additional vector and scalar fields, in strong-field regimes it cannot be assumed a priori that there is no coupling between matter and the additional fields. Rather, it is widely recognized in the physics community that, since the additional fields also constitute parts of the gravitational interaction, their contribution to the binding energy of compact objects may be nonnegligible, even if we assume that they are not directly coupled to a single test particle (see Refs.~\cite{Eardley:1975fgi,Will:1989sk,Zhang:2017sym,Higashino:2022izi,Quartin:2023tpl} and Refs.~\cite{Foster:2007gr,Lin:2018ken,Zhang:2019iim,Taherasghari:2022wfs,Taherasghari:2023rwn,Taherasghari:2025mlf} for scalar-tensor theories and vector-tensor theories). Note that, in this action~\eqref{action-all}, the vector field is by default defined as a covariant vector field $A_{\mu}$, and its contravariant counterpart is obtained by raising the index with the metric, i.e., $A^{\mu}\equiv g^{\mu\nu}A_{\nu}$. Unless otherwise specified, this definition is used throughout the paper.
\par
Varying the action~\eqref{action-all} with respect to the metric $g_{\mu\nu}$, the vector field $A_{\mu}$, and the scalar field $\phi$ separately yields the corresponding equations of motion
\begin{eqnarray}
  G^{\mu\nu} -S^{\mu\nu} &=& \frac{8\pi G}{c^4}T_{m}^{\mu\nu},\label{Eq1-20266162330}\\
  E^{\mu} &=& \frac{8\pi G}{c^4}J_{m}^{\mu},\label{Eq2-20266162330}\\
  E_{\phi} &=& \frac{8\pi G}{c^4}J_{\phi}. \label{Eq3-20266162330}
\end{eqnarray}
Here, the matter stress-energy tensor, the matter current, and the scalar source are defined as: $T_{m}^{\mu\nu}\equiv \tfrac{2}{\sqrt{-g}}\tfrac{\delta (\sqrt{-g}\mathcal{L}_{m})}{\delta g_{\mu\nu}}$, $J_{m}^{\mu}\equiv -\tfrac{1}{\sqrt{-g}}\tfrac{\delta (\sqrt{-g}\mathcal{L}_{m})}{\delta A_{\mu}}$, and $J_{\phi}\equiv -\tfrac{1}{\sqrt{-g}}\tfrac{\delta (\sqrt{-g}\mathcal{L}_{m})}{\delta \phi}$. On the left-hand side of the equations of motion, $G^{\mu\nu}$ is the Einstein tensor, and the various gravitational terms are given by $S^{\mu\nu}=\tfrac{1}{\sqrt{-g}}\tfrac{\delta (\sqrt{-g}\mathcal{L}_{g})}{\delta g_{\mu\nu}}+G^{\mu\nu}$, $E^{\mu}=\tfrac{1}{2 \sqrt{-g}}\tfrac{\delta (\sqrt{-g}\mathcal{L}_{g})}{\delta A_{\mu}}$, and $E_{\phi}=\tfrac{1}{2 \sqrt{-g}}\tfrac{\delta (\sqrt{-g}\mathcal{L}_{g})}{\delta \phi}$, respectively.
\par
We now use diffeomorphism invariance to derive the identity that generalizes the contracted Bianchi identity of GR. We consider an infinitesimal coordinate transformation $x^{\mu}\rightarrow x^{\mu}+\xi^{\mu}$. From the active point of view, the transformation rules for the fields are given as follows:
\begin{eqnarray}
  \delta_{\xi} g_{\mu\nu}(x)&=&g'_{\mu\nu}(x)-g_{\mu\nu}(x)=-\nabla_{\mu}\xi_{\nu}-\nabla_{\nu}\xi_{\mu},\label{InfiTrans1-2026772209}\\
  \delta_{\xi} A_{\mu}(x)&=& A'_{\mu}(x)-A_{\mu}(x)=-\xi^{\alpha}\nabla_{\alpha}A_{\mu}-A_{\alpha}\nabla_{\mu}\xi^{\alpha},\label{InfiTrans2-2026772209}\\
  \delta_{\xi} \phi (x) &=& \phi' (x)-\phi (x) = -\xi^{\alpha}\nabla_{\alpha}\phi . \label{InfiTrans3-2026772209}
\end{eqnarray}
Consequently, the variation of the action takes the form
\begin{eqnarray}
  \delta_{\xi}S_{g} &=& \frac{c^3}{16\pi G} \int d^4x\sqrt{-g}\left(\frac{\delta (\sqrt{-g}\mathcal{L}_{g})}{\sqrt{-g}\delta g_{\mu\nu}}\delta_{\xi} g_{\mu\nu} +\frac{\delta (\sqrt{-g}\mathcal{L}_{g})}{\sqrt{-g}\delta A_{\mu}}\delta_{\xi} A_{\mu} +\frac{\delta (\sqrt{-g}\mathcal{L}_{g})}{\sqrt{-g}\delta \phi}\delta_{\xi} \phi \right)\nonumber\\
  &=& -\frac{c^3}{8\pi G}\int d^4x\sqrt{-g}\xi_{\mu}\left[ \nabla_{\nu}(G^{\mu\nu}-S^{\mu\nu}-E^{\nu}A^{\mu})+E^{\nu}\nabla^{\mu}A_{\nu} +E_{\phi}\nabla^{\mu}\phi \right],\label{deltaS-20266291710}
\end{eqnarray}
where $S_{g}$ denotes the gravitational part in the action~\eqref{action-all}, i.e., $S_{g} = \tfrac{c^3}{16\pi G}\int d^{4}x\sqrt{-g} \mathcal{L}_{g}$. In view of the diffeomorphism invariance of the theory, the variation of the action is zero, i.e. $\delta_{\xi}S_{g}= 0$. Since the coordinate transformation parameter $\xi^{\mu}$ is arbitrary, it follows from Eq.~\eqref{deltaS-20266291710} that the diffeomorphism identity reads:
\begin{eqnarray}
  \nabla_{\nu}(G^{\mu\nu}-S^{\mu\nu}-E^{\nu}A^{\mu})+E^{\nu}\nabla^{\mu}A_{\nu} +E_{\phi}\nabla^{\mu}\phi\equiv 0.\label{identity-20266291909}
\end{eqnarray}
It is worth noting that this identity can also be obtained directly by combining the terms on the left-hand sides of the equations of motion~\eqref{Eq1-20266162330}-\eqref{Eq3-20266162330}. Furthermore, combining this identity with the equations of motion~\eqref{Eq1-20266162330}-\eqref{Eq3-20266162330}, we obtain the relation for the matter sector:
\begin{eqnarray}
  \nabla_{\nu}(T_{m}^{\mu\nu}-J_{m}^{\nu}A^{\mu}) +J_{m}^{\nu}\nabla^{\mu}A_{\nu} +J_{\phi}\nabla^{\mu}\phi = 0.\label{identityMatter-20266291909}
\end{eqnarray}
The results of Eqs.~\eqref{identity-20266291909} and~\eqref{identityMatter-20266291909} are consistent with those in Ref.~\cite{Gralla:2013rwa}. Thus far, it is readily seen that, in modified gravity~\eqref{action-all}, the source stress-energy $T^{\mu\nu}_{m}$ is not conserved in general, which is in marked contrast to the conservation of the source stress-energy tensor in GR. In particular, for minimally coupled matter with $J_m^\mu=J_\phi=0$, the usual relation $\nabla_\nu T_m^{\mu\nu}=0$ is recovered.
\par
The choices $A_\mu$ and $A^\mu$ describe the same theory. However, their metric variations differ because different vector components are held fixed. The above conclusions~\eqref{identity-20266291909} and~\eqref{identityMatter-20266291909} are based on the definition of the action $S\equiv S\left[ g_{\mu\nu},A_{\mu},\phi,\psi_{m} \right]$. If we adopt an alternative definition, namely $S\equiv S\left[ g_{\mu\nu},A^{\mu},\phi,\psi_{m} \right]$ with $A_\mu\equiv g_{\mu\nu}A^\nu$, then the forms of both the identity and the matter relation will be different. Following the same procedure as in Eqs.~\eqref{action-all}-\eqref{identityMatter-20266291909}, we obtain
\begin{eqnarray}
  \nabla_{\nu}(G^{\mu\nu}-S'^{\mu\nu}+E^{\mu}A^{\nu})+E_{\nu}\nabla^{\mu}A^{\nu}+E_{\phi}\nabla^{\mu}\phi &\equiv& 0, \label{identity1-20266291909} \\
  \nabla_{\nu}(T_{m}'^{\mu\nu}+J_{m}^{\mu}A^{\nu})+J_{m\;\nu}\nabla^{\mu}A^{\nu}+J_{\phi}\nabla^{\mu}\phi &=& 0, \label{identityMatter1-20266291909}
\end{eqnarray}
where $S'^{\mu\nu}$ and $T_{m}'^{\mu\nu}$ are the terms derived from definition $S\equiv S\left[ g_{\mu\nu},A^{\mu},\phi,\psi_{m} \right]$. They are defined in the same manner as $S^{\mu\nu}$ and $T_{m}^{\mu\nu}$, but their explicit forms generally differ from those because the fundamental definition of the vector field is different. By the chain rule, the relations between $S'^{\mu\nu}$ and $S^{\mu\nu}$ and between $T_{m}'^{\mu\nu}$ and $T_{m}^{\mu\nu}$ are obtained as follows:
\begin{eqnarray}
  S'^{\mu\nu} &\equiv& G^{\mu\nu}+\frac{1}{\sqrt{-g}}\left( \frac{\delta (\sqrt{-g}\mathcal{L}_{g})}{\delta g_{\mu\nu}} +\frac{\delta (\sqrt{-g}\mathcal{L}_{g})}{\delta A_{\alpha}}\frac{\delta A_{\alpha}}{\delta g_{\mu\nu}} \right) = S^{\mu\nu} +2 E^{(\mu}A^{\nu)}, \label{GWseT0-20267221052}
  \\
  T_{m}'^{\mu\nu} &\equiv& \frac{2}{\sqrt{-g}}\left( \frac{\delta (\sqrt{-g}\mathcal{L}_{m})}{\delta g_{\mu\nu}} +\frac{\delta (\sqrt{-g}\mathcal{L}_{m})}{\delta A_{\alpha}}\frac{\delta A_{\alpha}}{\delta g_{\mu\nu}} \right) = T_{m}^{\mu\nu} -2 J_{m}^{(\mu}A^{\nu)}.
\end{eqnarray}
Here, $K^{(\mu\nu)}\equiv \tfrac{1}{2}(K^{\mu\nu}+K^{\nu\mu})$ where $K$ denotes an arbitrary quantity.
\par
Equations~\eqref{identity-20266291909} and~\eqref{identity1-20266291909} are identities, whereas Eqs.~\eqref{identityMatter-20266291909} and~\eqref{identityMatter1-20266291909} express matter balance on solutions of the coupled field equations. We now expand about a vacuum solution with a Minkowski metric and constant vector and scalar fields in the Minkowski background. Applying identities~\eqref{identity-20266291909} and~\eqref{identity1-20266291909} to both the rapidly varying perturbations and the slow response, we obtain the following relations:
\begin{eqnarray}
  \text{For }S\left[ A_{\mu},\cdots \right]:&&  \partial_{\mu}Q^{(1)\mu\nu}=0,\quad \partial_{\mu}\bar{Q}^{(1)\mu\nu}=0, \quad \left(Q^{\mu\nu}\equiv G^{\mu\nu}-S^{\mu\nu}-E^{\mu}\bar{A}^{(0)\nu}\right) \label{identityQ1-20261021910}\\
  \text{For }S\left[ A^{\mu},\cdots \right]:&&  \partial_{\mu}Q'^{(1)\mu\nu}=0,\quad \partial_{\mu}\bar{Q}'^{(1)\mu\nu}=0,\quad  \left(Q'^{\mu\nu}\equiv G^{\mu\nu}-S'^{\mu\nu}+E^{\nu}\bar{A}^{(0)\mu}\right) \label{identityQ2-20261021910}
\end{eqnarray}
where $S\left[ A_{\mu},\cdots \right]$ and $S\left[ A^{\mu},\cdots \right]$ abbreviate the two choices of fundamental vector variable. From relations~\eqref{identityQ1-20261021910} and~\eqref{identityQ2-20261021910} and the equations of motion~\eqref{Eq1-20266162330}-\eqref{Eq3-20266162330}, the linear field equations incorporating the backreaction on the spacetime background and a conserved source are obtained:
\begin{eqnarray}
  \text{For }S\left[ A_{\mu},\cdots \right]:&& Q^{(1)\mu\nu}+\bar{Q}^{(1)\mu\nu} =\frac{8\pi G}{c^4}\left( T_{m}^{\mu\nu}-J_{m}^{\mu}\bar{A}^{(0)\nu} -\frac{c^4}{8\pi G} \sum_{\ell_{1}+\ell_{2}\ge 2}Q^{(\ell_{1},\ell_{2})\mu\nu}[\bar{\epsilon},\epsilon] \right), \label{identityQ3-20261021910}
  \\
  \text{For }S\left[ A^{\mu},\cdots \right]:&& Q'^{(1)\mu\nu}+\bar{Q}'^{(1)\mu\nu} =\frac{8\pi G}{c^4}\left( T_{m}'^{\mu\nu}+J_{m}^{\nu}\bar{A}^{(0)\mu} -\frac{c^4}{8\pi G} \sum_{\ell_{1}+\ell_{2}\ge 2}Q'^{(\ell_{1},\ell_{2})\mu\nu}[\bar{\epsilon},\epsilon] \right).\label{identityQ4-20261021910}
\end{eqnarray}
The linear identities~\eqref{identityQ1-20261021910} and~\eqref{identityQ2-20261021910} ensure that the complete right-hand sides of these equations have vanishing ordinary divergence. Their nonlinear field terms describe both sources of the radiation and the backreaction on the slow background. Short-wavelength averaging selects the latter contribution (see Sec.~\ref{IPandQoM-2026761755}),
\begin{eqnarray}
  \text{For }S\left[ A_{\mu},\cdots \right]:&&  \tau^{\mu\nu} \equiv -\frac{c^4}{8\pi G}\left\langle  \sum_{\ell_{1}+\ell_{2}\ge 2}Q^{(\ell_{1},\ell_{2})\mu\nu}[\bar{\epsilon},\epsilon]  \right\rangle\nonumber\\
  &&\qquad = \frac{c^4}{8\pi G}\left\langle  S^{(2)\mu\nu}-G^{(2)\mu\nu}+E^{(2)\mu}\bar{A}^{(0)\nu} +\mathcal{O}(\epsilon^3) \right\rangle,\label{GWseT1-20267221052}\\
  \text{For }S\left[ A^{\mu},\cdots \right]:&&  \tau^{\mu\nu} \equiv -\frac{c^4}{8\pi G}\left\langle  \sum_{\ell_{1}+\ell_{2}\ge 2}Q'^{(\ell_{1},\ell_{2})\mu\nu}[\bar{\epsilon},\epsilon] \right\rangle\nonumber\\
  &&\qquad = \frac{c^4}{8\pi G}\left\langle  S'^{(2)\mu\nu}-G^{(2)\mu\nu}-E^{(2)\nu}\bar{A}^{(0)\mu} +\mathcal{O}(\epsilon^3) \right\rangle.\label{GWseT2-20267221052}
\end{eqnarray}
In the flat vacuum wave zone, Eqs.~\eqref{identityQ1-20261021910} and~\eqref{identityQ2-20261021910} imply $\partial_\mu\tau^{\mu\nu}=0$. In the next section, we shall show from different perspectives that $\tau^{\mu\nu}$ at the retained order is the GW stress-energy tensor. The tensor includes the quadratic vector-equation term required by the diffeomorphism identity and is generally nonsymmetric. With $x^0=ct$, the energy balance uses $\tau^{00}$ as the energy density and $c\tau^{i0}$ as the energy flux density. The quantity $\tau^{0i}$ is not interchangeable with $\tau^{i0}$. 
\par
Furthermore, combining Eqs.~\eqref{identityQ1-20261021910}-\eqref{identityQ4-20261021910} gives the matter-field exchange equations:
\begin{eqnarray}
  \text{For }S\left[ A_{\mu},\cdots \right]:&& \partial_{\mu}\left( T_{m}^{\mu\nu}-J_{m}^{\mu}\bar{A}^{(0)\nu} \right)= \frac{c^4}{8\pi G}\partial_{\mu}\left(  G^{(2)\mu\nu}-S^{(2)\mu\nu}-E^{(2)\mu}\bar{A}^{(0)\nu} +\mathcal{O}(\epsilon^3) \right), \label{matter1-20261022142}
  \\
  \text{For }S\left[ A^{\mu},\cdots \right]:&& \partial_{\mu}\left( T_{m}'^{\mu\nu}+J_{m}^{\nu}\bar{A}^{(0)\mu} \right)= \frac{c^4}{8\pi G}\partial_{\mu}\left(  G^{(2)\mu\nu}-S'^{(2)\mu\nu}+E^{(2)\nu}\bar{A}^{(0)\mu} +\mathcal{O}(\epsilon^3) \right).\label{matter2-20261022142}
\end{eqnarray}
Equations~\eqref{matter1-20261022142} and~\eqref{matter2-20261022142} show that the matter part of the source does not separately satisfy an ordinary divergence conservation law: it exchanges energy and momentum with the nonlinear gravitational field. The complete source, including the latter terms, is conserved. This distinction is essential for relating source dynamics to the outgoing radiation flux.
\par
These balance relations identify the conserved source needed to connect radiation to source evolution. The symmetric quadratic term in the metric equation is often treated as the conventional GW stress-energy tensor~\cite{Saffer:2017ywl,Heisenberg:2025roe}. However, when a background vector is present, it need not necessarily serve as that source. We examine this distinction in Sec.~\ref{SecComparison-20269232159}. We next complete the four standard constructions with the same diffeomorphism identity and compare their averaged tensors with Eqs.~\eqref{GWseT1-20267221052} and~\eqref{GWseT2-20267221052}.

\section{GW stress-energy tensor}\label{GWSET-20269251143}
We now investigate the GW stress-energy (pseudo)tensor within the wave zone in the Minkowski background. In the GW community, two complementary perspectives are commonly adopted to describe the nature of gravity~\cite{Maggiore2007}: the geometrical perspective and the field-theoretical perspective. These two perspectives naturally lead to different methods to define the GW stress-energy (pseudo)tensor. The commonly used methods in GR include: the perturbed field equation method~\cite{Isaacson:1968hbi,Isaacson:1968zza}, the Landau-Lifshitz method~\cite{landau1975classical}, the perturbed action method~\cite{Isaacson:1968hbi,Isaacson:1968zza,Saffer:2017ywl}, and the Noether current method~\cite{Noether:1918zz}. They provide different definitions of the GW stress-energy (pseudo)tensor, but all lead to the same result. In this paper, we will discuss these four methods and extend them to provide a definition of the conserved GW stress-energy (pseudo)tensor that is applicable to general metric theories of gravity. Here, we use $S\left[ A_{\mu},\cdots \right]$ as an example for our analysis. The results for the candidate definition $S\left[ A^{\mu},\cdots \right]$ are presented in Appendix~\ref{appendix-GWSET-20269251143}.

\subsection{The field-theoretical perspective}
For the field-theoretical perspective, GWs are regarded as fields propagating on a fixed background spacetime. At this point, the slowly varying fields $\{\delta \bar{g}_{\mu\nu},\delta \bar{A}_{\mu},\delta \bar{\phi}\}$ are treated as higher-order contributions to the gravitational perturbation and, together with the high-frequency parts, are contained in the perturbations about the background. Therefore, we first hold the reference background fixed and take:
\begin{eqnarray}
  \bar{g}_{\mu\nu}\approx \eta_{\mu\nu},\quad \bar{A}_{\mu}\approx (\bar{A},0,0,0), \quad \bar{\phi}\approx \bar{\phi}^{(0)}. \label{background31-2026761811}
\end{eqnarray}
Here, $\eta_{\mu\nu}$ is the Minkowski metric, and $\bar{A}$ and $\bar{\phi}^{(0)}$ are constants. For the perturbations defined in Eqs.~\eqref{perturbation1-2026761850}-\eqref{perturbation3-2026761850}, the quadratic gravitational action is
\begin{eqnarray}
  S_{g}^{(2)}\left[ h_{\mu\nu},\delta A_{\mu},\delta\phi \right]=\int dtd^{3}\vec{x} \;\tilde{\mathcal{L}}_{g}\left[ h_{\mu\nu},\delta A_{\mu},\delta\phi \right],\label{Sg2-2026771258}
\end{eqnarray}
where $\tilde{\mathcal{L}}_{g}\equiv \tfrac{c^4}{16\pi G}\left(\sqrt{-g}\mathcal{L}_{g}\right)^{(2)}$. Based on this action, we can define the GW stress-energy tensor via the Noether current method and the perturbed action method, respectively.
\par
\textit{\textbf{The Noether current method.}} In the Minkowski spacetime, Noether's theorem states that, for each global symmetry  transformation, there is a current which, when evaluated on a classical solution of the equations of motion, is conserved~\cite{Maggiore2007,Noether:1918zz}. In particular, the spacetime translation symmetry leads to the conservation of the stress-energy tensor. For the second-order perturbed action~\eqref{Sg2-2026771258}, under spacetime translation transformations, one obtains a conserved current $j^{\mu}_{\;\;\nu}$ which satisfies $\partial_{\mu}j^{\mu}_{\;\;\nu}=0$~\cite{Noether:1918zz,Olver1993,Iyer:1994ys,Maggiore2007,Saffer:2017ywl}, and its specific form is
\begin{eqnarray}
  j^{\mu}_{\;\;\nu}&=& \delta^{\mu}_{\;\;\nu}\tilde{\mathcal{L}}_{g}-\sum^{s}_{n=1}\sum^{n-1}_{\ell=0}(-1)^{\ell}\left( \partial_{\alpha_{1}}\cdots \partial_{\alpha_{\ell}} \frac{\partial\tilde{\mathcal{L}}_{g}}{\partial(\partial_{\mu}\partial_{\alpha_{1}}\cdots \partial_{\alpha_{n-1}}\phi^{A})} \right)\partial_{\alpha_{\ell+1}}\cdots\partial_{\alpha_{n-1}}\partial_{\nu}\phi^{A} \nonumber\\
  &=& \delta^{\mu}_{\;\;\nu}\tilde{\mathcal{L}}_{g}-\frac{\partial\tilde{\mathcal{L}}_{g}}{\partial(\partial_{\mu}h_{\alpha\rho})}\partial_{\nu}h_{\alpha\rho} -\frac{\partial\tilde{\mathcal{L}}_{g}}{\partial(\partial_{\mu}\delta A_{\alpha})}\partial_{\nu}\delta A_{\alpha} -\frac{\partial\tilde{\mathcal{L}}_{g}}{\partial(\partial_{\mu}\delta\phi)}\partial_{\nu}\delta\phi +\cdots, \label{jCSE-2026771442}
\end{eqnarray}
where the index $A$ in $\phi^A$ collectively labels the perturbation fields and their tensor indices, and the ellipsis in the second line denotes contributions from derivatives beyond first order. Here, $s$ denotes the highest derivative order of the fields appearing in $\tilde{\mathcal{L}}_g$, and any empty product of derivative operators is understood as the identity operator. For example, $\partial_{\alpha_1}\cdots\partial_{\alpha_\ell}=1$ when $\ell=0$. The current $j^{\mu}_{\;\;\nu}$ is known as the canonical stress-energy tensor. Since the energy of GWs cannot be well-defined at a local point in space, the GW stress-energy tensor is typically assumed to be the short-wavelength average of the conserved current,
\begin{eqnarray}
  \tau^{\mu\nu}=\langle j^{\mu\nu} \rangle,\label{GWSET-N-2026771443}
\end{eqnarray}
Following a similar procedure, we obtain the GW stress-energy tensor based on $S\left[ A^{\mu},\cdots \right]$, as given in Eq.~\eqref{GWSET-N1-2026771443} of the Appendix. For $s=1$, Eq.~\eqref{jCSE-2026771442} displays the metric, vector, and scalar contributions explicitly. The averaged tensor is generally nonsymmetric and satisfies $\partial_\mu\tau^{\mu\nu}=0$. A constant preferred vector preserves translation invariance but need not preserve Lorentz invariance of the perturbation theory. A conserved translation current therefore need not admit the usual Lorentz-based Belinfante symmetrization.
\par
\textit{\textbf{The perturbed action method.}} From the geometrical perspective, this method is equivalent to the perturbed field equation method, and it follows immediately from recasting the results of the latter into the variational form of the action~\cite{Dong:2024zal}. To avoid redundancy, we shall give a field-theoretical analysis to obtain the GW stress-energy tensor. In this method, $h_{\mu\nu}$, $\delta A_{\mu}$, and $\delta\phi$ in the action~\eqref{Sg2-2026771258} can be regarded as ``matter'' fields, which serve as the sources of the background gravitational fields. Therefore, in analogy with the conventional definition of the matter stress-energy tensor, the GW stress-energy tensor is obtained by varying the action with respect to the background gravitational fields. In GR, the GW stress-energy tensor can be obtained by varying the action with respect to the background metric~\cite{Saffer:2017ywl}. However, in modified gravity, the presence of the added fields causes variation with respect to the metric alone to give only one part of the background response, see Eqs.~\eqref{identity-20266291909} and~\eqref{identity1-20266291909} in Sec.~\ref{Theoretical-Framework-2026772109}. Consequently, it is reasonable to expect that the method needs to be generalized. The conserved tensor is selected by the diffeomorphism identity linking variations of all background fields, so the vector and scalar variations must be included before specializing to a constant background. Therefore, we apply the diffeomorphism variations~\eqref{InfiTrans1-2026772209}-\eqref{InfiTrans3-2026772209} to the background fields in $S_g^{(2)}$,
\begin{eqnarray}
  \delta_{\xi}S_{g}^{(2)} &=& \frac{1}{c}\int d^4x\sqrt{-\bar{g}}\left[\frac{\delta \tilde{\mathcal{L}}_{g}}{\sqrt{-\bar{g}}\delta \bar{g}_{\mu\nu}}\delta_{\xi} \bar{g}_{\mu\nu} +\frac{\delta \tilde{\mathcal{L}}_{g}}{\sqrt{-\bar{g}}\delta \bar{A}_{\mu}}\delta_{\xi} \bar{A}_{\mu} +\frac{\delta \tilde{\mathcal{L}}_{g}}{\sqrt{-\bar{g}}\delta \bar{\phi}}\delta_{\xi} \bar{\phi} \right]\nonumber\\
  &=& \frac{1}{c}\int d^4x\sqrt{-\bar{g}}\xi_{\nu}\partial_{\mu}\left[ \frac{2}{\sqrt{-\bar{g}}}\frac{\delta \tilde{\mathcal{L}}_{g}}{\delta \bar{g}_{\mu\nu}} +\frac{1}{\sqrt{-\bar{g}}}\frac{\delta \tilde{\mathcal{L}}_{g}}{\delta \bar{A}_{\mu}}\bar{A}^{\nu} \right].\label{PerAction1-2026772258}
\end{eqnarray}
Here, we have employed the on-shell conditions, which cause the linear terms involving $\delta_{\xi} h_{\mu\nu}$, $\delta_{\xi} \delta A_{\mu}$, and $\delta_{\xi} \delta\phi$ on the right-hand side of the first equality to vanish. The background variation must be performed before imposing its flat, constant values. The last line of Eq.~\eqref{PerAction1-2026772258} uses $\partial_\alpha\bar A_\mu=0$ and $\partial_\alpha\bar\phi=0$. The scalar variation contributes no gradient term in this limit, although scalar waves still contribute to the quadratic action and tensor. Diffeomorphism invariance and the arbitrariness of $\xi_\nu$ then imply that the bracket in that equation has vanishing first-index divergence. Therefore, after short-wavelength averaging, we define the GW stress-energy tensor as
\begin{eqnarray}
  \tau^{\mu\nu}= \left\langle \frac{2}{\sqrt{-\bar{g}}}\frac{\delta \tilde{\mathcal{L}}_{g}}{\delta \bar{g}_{\mu\nu}} +\frac{1}{\sqrt{-\bar{g}}}\frac{\delta \tilde{\mathcal{L}}_{g}}{\delta \bar{A}_{\mu}}\bar{A}^{\nu} \right\rangle.\label{GWSET-A-2026771443} 
\end{eqnarray}
For the action $S\left[ A^{\mu},\cdots \right]$, the corresponding GW stress-energy tensor is \eqref{GWSET-A1-2026771443} in the Appendix. As in the Noether construction, the resulting GW stress-energy tensor is generally nonsymmetric and conserved on its first index.

\subsection{The geometrical perspective}
For the geometrical perspective, an important concept is the Isaacson picture, see Sec.~\ref{IPandQoM-2026761755}. At this point, since the GW stress-energy (pseudo)tensor is defined by its backreaction on the background spacetime, we can no longer a priori approximate the background spacetime as fixed. Rather, the background spacetime should be treated as more general, and the slowly varying components can be isolated as in Eqs.~\eqref{background1-2026761811}-\eqref{background3-2026761811}, while keeping the reference vacuum as
\begin{eqnarray}
  \bar{g}^{(0)}_{\mu\nu}=\eta_{\mu\nu},\quad \bar{A}^{(0)}_{\mu}=(\bar{A},0,0,0),\quad \bar{\phi}^{(0)}=\text{const.}.\label{BackgroundValue-20269221651}
\end{eqnarray}
Therefore, $\{\delta \bar{g}_{\mu\nu},\delta \bar{A}_{\mu},\delta \bar{\phi}\}$ represent slowly varying components of the background arising from the backreaction of high-frequency GWs $\{h_{\mu\nu},\delta A_{\mu},\delta\phi\}$. Based on the full equations of motion~\eqref{Eq1-20266162330}-\eqref{Eq3-20266162330}, using the linear wave equations, and retaining terms of backreaction order, we can obtain the linearized and sourced equations of motion for $\{\delta \bar{g}_{\mu\nu},\delta \bar{A}_{\mu},\delta \bar{\phi}\}$,
\begin{eqnarray}
  \bar{G}^{(1)\mu\nu} -\bar{S}^{(1)\mu\nu} &=& \frac{8\pi G}{c^4}\left[  T_{m}^{(2)\mu\nu} +\tfrac{c^4}{8\pi G}\big( S^{(2)\mu\nu}-G^{(2)\mu\nu} \big) \right],\label{BGEq1-2026781504}\\
  \bar{E}^{(1)\mu} &=& \frac{8\pi G}{c^4} \left[ J_{m}^{(2)\mu} -\tfrac{c^4}{8\pi G}E^{(2)\mu} \right],\label{BGEq2-2026781504}\\
  \bar{E}^{(1)}_{\phi} &=& \frac{8\pi G}{c^4} \left[ J_{\phi}^{(2)} -\tfrac{c^4}{8\pi G}E^{(2)}_{\phi} \right].\label{BGEq3-2026781504}
\end{eqnarray}
It is readily seen from these three equations that $G^{(2)\mu\nu}$, $S^{(2)\mu\nu}$, $E^{(2)\mu}$, and $E^{(2)}_{\phi}$ serve as sources for the slowly varying components of the background $\{\delta \bar{g}_{\mu\nu},\delta \bar{A}_{\mu},\delta \bar{\phi}\}$.
\par
\textit{\textbf{The perturbed field equation method.}} In this method, the quadratic terms of the perturbations $\{h_{\mu\nu},\delta A_{\mu},\delta\phi\}$ in the equations of motion~\eqref{BGEq1-2026781504}-\eqref{BGEq3-2026781504} serve as the sources for the slowly varying components of the background $\{\delta \bar{g}_{\mu\nu},\delta \bar{A}_{\mu},\delta \bar{\phi}\}$. The metric equation identifies the quadratic source $t^{\mu\nu}=c^4(S^{(2)\mu\nu}-G^{(2)\mu\nu})/(8\pi G)$. It represents only the metric part of the backreaction. The vector and scalar equations also have quadratic sources, and the diffeomorphism identity~\eqref{identity-20266291909} determines how they enter the energy-momentum balance. When the linear vacuum equations hold on a constant background, they give
\begin{eqnarray}
  \partial_{\mu}\left( t^{\mu\nu} +\tfrac{c^4}{8\pi G}E^{(2)\mu}\bar{A}^{(0)\nu} \right)=0.
\end{eqnarray}
Averaging this expression gives the conserved source for the slow response. Therefore, the GW stress-energy tensor is defined as
\begin{eqnarray}
  \tau^{\mu\nu}=\frac{c^4}{8\pi G}\left\langle S^{(2)\mu\nu}-G^{(2)\mu\nu}+E^{(2)\mu}\bar{A}^{(0)\nu} \right\rangle. \label{perField-20267232110} 
\end{eqnarray}
We also derive the GW stress-energy tensor from $S\left[A^{\mu},\cdots\right]$ and present the result in Eq.~\eqref{perField1-20267232110} of the Appendix. Similar to the GW stress-energy tensors obtained by the previous two methods, these stress-energy tensors are also nonsymmetric, and only their first index satisfies the conservation equation.
\par
\textit{\textbf{The Landau-Lifshitz method.}} This approach constructs a gravitational pseudotensor such that the total sum of the matter terms and the gravitational terms satisfies an ordinary conservation law, thereby yielding, in the far-field approximation, a coordinate-dependent yet physically meaningful GW stress-energy pseudotensor~\cite{landau1975classical}. In the Landau-Lifshitz method, a 4-tensor density is constructed~\cite{landau1975classical,Saffer:2017ywl},
\begin{eqnarray}
  \mathcal{H}^{\alpha\mu\beta\nu}\equiv \mathfrak{g}^{\alpha\beta}\mathfrak{g}^{\mu\nu}-\mathfrak{g}^{\alpha\nu}\mathfrak{g}^{\beta\mu}, \label{H-20267141515}
\end{eqnarray}
where $\mathfrak{g}^{\mu\nu}\equiv \sqrt{-g}g^{\mu\nu}$. It has the remarkable property that
\begin{eqnarray}
  \partial_{\mu}\partial_{\nu}\mathcal{H}^{\alpha\mu\beta\nu}=2(-g) G^{\alpha\beta}+\frac{16\pi G}{c^4}(-g)t_{LL}^{\alpha\beta}.\label{ddH-20267232102}
\end{eqnarray}
Here $G^{\alpha\beta}$ is the Einstein tensor and $t_{LL}^{\alpha\beta}$ is the Landau-Lifshitz pseudotensor. The specific form of $(-g)t_{LL}^{\alpha\beta}$ is~\cite{landau1975classical,Saffer:2017ywl}
\begin{eqnarray}
  (-g)t_{LL}^{\alpha\beta} &=& \frac{c^4}{16\pi G}\Big[ 2\partial_{\mu}\mathfrak{g}^{\alpha[\beta}\partial_{\nu}\mathfrak{g}^{\mu]\nu} +\frac{1}{2}g^{\alpha\beta}g_{\mu\nu}\partial_{\rho}\mathfrak{g}^{\mu\gamma}\partial_{\gamma}\mathfrak{g}^{\rho\nu} -2g^{\mu(\alpha}g_{\rho\gamma}\partial_{\nu}\mathfrak{g}^{\beta)\gamma}\partial_{\mu}\mathfrak{g}^{\rho\nu} \nonumber\\
  && +g_{\rho\gamma}g^{\mu\nu}\partial_{\mu}\mathfrak{g}^{\alpha\rho}\partial_{\nu}\mathfrak{g}^{\beta\gamma} +\frac{1}{8}\left( 2 g^{\alpha\mu}g^{\beta\nu} -g^{\alpha\beta}g^{\mu\nu} \right)\left( 2 g_{\rho\gamma}g_{\lambda\sigma} -g_{\gamma\lambda}g_{\rho\sigma} \right)\partial_{\mu}\mathfrak{g}^{\rho\sigma}\partial_{\nu}\mathfrak{g}^{\gamma\lambda}  \Big].
\end{eqnarray}
Antisymmetry of $\mathcal H^{\alpha\mu\beta\nu}$ under $\alpha\leftrightarrow\mu$ implies $\partial_\alpha\partial_\mu\partial_\nu\mathcal H^{\alpha\mu\beta\nu}=0$. In GR, the field equations $G^{\mu\nu}=8\pi G T^{\mu\nu}/c^4$ turn this identity into the usual matter-plus-gravity conservation law, and its averaged quadratic vacuum contribution $\left\langle \left( (-g)t_{LL}^{\alpha\beta} \right)^{(2)} \right\rangle$ defines the GW stress-energy pseudotensor~\cite{landau1975classical,Saffer:2017ywl}. In the present theory, Eqs.~\eqref{Eq1-20266162330}-\eqref{Eq3-20266162330} introduce additional field contributions. The complete Landau-Lifshitz density is conserved before the separation into waves and background. Extracting its radiative part requires retaining the slow background response and its vector equation at the same order as the quadratic wave terms. The constant zeroth-order superpotential contributes no derivatives, and the linear high-frequency term averages to zero. At the leading backreaction order, the remaining identity is
\begin{eqnarray}
  0 &=& \left(\partial_{\alpha}\partial_{\mu}\partial_{\nu}\mathcal{H}^{\alpha\mu\beta\nu}\right)^{(2)} \nonumber\\
  &=& 2\partial_{\alpha}\left( (-g)G^{\alpha\beta} \right)^{(2)} +\frac{16\pi G}{c^4}\partial_{\alpha}\left( (-g)t_{LL}^{\alpha\beta} \right)^{(2)} \nonumber\\
  &=& \frac{16\pi G}{c^4}\partial_{\alpha}\left( (-g)T_{m}^{\alpha\beta} +(-g)t_{LL}^{\alpha\beta} +\tfrac{c^4}{8\pi G} (-g)S^{\alpha\beta} \right)^{(2)} +2\partial_{\alpha}\left( (-\bar{g})\bar{S}^{\alpha\beta} - (-\bar{g})\bar{G}^{\alpha\beta} \right)^{(1)}\nonumber\\
  &=& \frac{16\pi G}{c^4}\partial_{\alpha}\left[ (-g) \left( T_{m}^{\alpha\beta} -J_{m}^{\alpha}A^{\beta} \right) +(-g) \left( t_{LL}^{\alpha\beta} +\tfrac{c^4}{8\pi G}S^{\alpha\beta} +\tfrac{c^4}{8\pi G}E^{\alpha}A^{\beta} \right) \right]^{(2)}.\label{HLL-20267142146}
\end{eqnarray}
Here, the third line uses the metric response equation~\eqref{BGEq1-2026781504}, and the last line additionally uses the diffeomorphism identity~\eqref{identity-20266291909} and the vector response equation~\eqref{BGEq2-2026781504}. This rearrangement isolates a completed matter contribution and a completed gravitational contribution with vanishing total ordinary divergence. In the vacuum wave zone, their averaged quadratic gravitational part defines 
\begin{eqnarray}
  \tau^{\mu\nu}=\left\langle \left[(-g) \left( t_{LL}^{\mu\nu} +\tfrac{c^4}{8\pi G}S^{\mu\nu} +\tfrac{c^4}{8\pi G}E^{\mu}A^{\nu} \right) \right]^{(2)} \right\rangle.
\end{eqnarray}
Equation~\eqref{HLL-20267142146} establishes first-index conservation in the vacuum wave zone, while the vector completion makes the radiative current generally nonsymmetric. Furthermore, combining Eq.~\eqref{ddH-20267232102}, $(-g)t_{LL}^{\mu\nu}$ and $-\tfrac{c^4}{8\pi G}(-g)G^{\mu\nu}$ differ only by a total derivative term, which, under the averaging operation, reduces to a boundary term and can therefore be neglected. Therefore, the GW stress-energy pseudotensor can be rewritten in the tensor form as
\begin{eqnarray}
  \tau^{\mu\nu}=\frac{c^4}{8\pi G}\left\langle S^{(2)\mu\nu}-G^{(2)\mu\nu}+E^{(2)\mu}\bar{A}^{(0)\nu} \right\rangle, \label{LLSE-20269212309}
\end{eqnarray}
Here, we have employed the linear vacuum equations. Therefore, the definition of the GW stress-energy tensors obtained by the Landau-Lifshitz method is identical to that given by the perturbed field equation method~\eqref{perField-20267232110} in the flat vacuum wave zone and at leading order in short-wavelength averaging. See also Eq.~\eqref{GWSE-202610191208} for the result based on $S\left[A^{\mu},\cdots\right]$.
\par
In a Minkowski background, all four methods described above yield the GW stress-energy tensors that satisfy the ordinary divergence conservation condition. Even though the definitions given by these methods differ, the final results are nevertheless consistent, which will be examined in the next section by taking Einstein-\ae ther theory as an example. In particular, when a non-Minkowski background is considered, the background is generally no longer constant. For the Noether current method, this implies that under spacetime translations the background quantities must also transform, and hence Eq.~\eqref{jCSE-2026771442} needs to be modified. For the perturbed action method, the perturbed field equation method, and the Landau-Lifshitz method, under diffeomorphisms the expressions no longer reduce to total ordinary derivatives, see Eq.~\eqref{identity-20266291909}. Consequently, the GW stress-energy tensors will no longer satisfy the ordinary-derivative conservation law obtained from the four methods above when a non-Minkowski background is considered.

\subsection{Comparison with conventional GW stress-energy tensor}\label{SecComparison-20269232159}
In general metric theories, the GW stress-energy tensor derived from our definition is conserved but not symmetric in its indices, whereas the conventional GW stress-energy tensor, constructed solely from the metric source, is symmetric in its indices but not conserved. Foster~\cite{Foster:2005fr,Foster:2007gr} adopted the Noether charge method in studying Einstein-\ae ther theory, which is based on the same principle as the Noether current method. The GW energy in our definition is equivalent to his, but Foster did not provide the stress-energy tensor of GWs. The GW stress-energy tensor defined in Refs.~\cite{Saffer:2017ywl,Heisenberg:2025roe} is called the conventional GW stress-energy tensor in this paper. In this subsection, we restrict ourselves to the Minkowski background and discuss the similarities and differences between the GW stress-energy tensor defined above and the conventional GW stress-energy tensor.
\par
The conventional GW stress-energy tensor is obtained from a direct extension of GR, namely, by varying the second-order perturbed action with respect to the background metric or by perturbing the metric field equations to second order:
\begin{eqnarray}
  \text{For }S\left[ A_{\mu},\cdots \right]:&\quad& t^{\mu\nu}\equiv\frac{c^4}{8\pi G}\left\langle S^{(2)\mu\nu}-G^{(2)\mu\nu} \right\rangle,\label{conventionalSET-20269211800}
  \\
  \text{For }S\left[ A^{\mu},\cdots \right]:&\quad& t'^{\mu\nu}\equiv\frac{c^4}{8\pi G}\left\langle S'^{(2)\mu\nu}-G^{(2)\mu\nu} \right\rangle.
\end{eqnarray}
Without loss of generality, we focus on $S\left[ A_{\mu},\cdots \right]$ in the following analysis. For pure metric theories or scalar-tensor theories, these two definitions, together with the Landau-Lifshitz method and the Noether current method, all yield equivalent GW stress-energy tensors. This conventional GW stress-energy tensor is conserved and symmetric in its indices. The GW stress-energy tensor obtained from our extension $\tau^{\mu\nu}$~\eqref{GWseT1-20267221052} is consistent with the one in Eq.~\eqref{conventionalSET-20269211800}. However, for vector-tensor theories with a nonzero background vector, the GW stress-energy tensor~\eqref{GWseT1-20267221052} defined in this paper begins to differ from the conventional GW stress-energy tensor~\eqref{conventionalSET-20269211800}.
\par
First, there is ambiguity in the definition of the conventional GW stress-energy tensor. The definitions of the conventional GW stress-energy tensor by the perturbed field equation method and the perturbed action method both lead to the tensor~\eqref{conventionalSET-20269211800}, see Ref.~\cite{Saffer:2017ywl}. However, the Noether current method and the Landau-Lifshitz method generally do not yield an GW stress-energy tensor equivalent to that in Eq.~\eqref{conventionalSET-20269211800}. For the Noether current method, (see the preceding subsection and Ref.~\cite{Saffer:2017ywl}), the GW stress-energy tensor that it usually yields is equivalent to that in Eq.~\eqref{GWseT1-20267221052}. However, Eq.~\eqref{GWseT1-20267221052} and Eq.~\eqref{conventionalSET-20269211800} are clearly not equivalent, differing by a term $E^{(2)\mu}\bar{A}^{(0)\nu}$. For the Landau-Lifshitz method, the resulting GW stress-energy tensor is also equivalent to that given in Eq.~\eqref{GWseT1-20267221052}, see the preceding subsection. In Ref.~\cite{Saffer:2017ywl}, the authors neglected the backreaction of gravity on the background when considering second-order perturbations of the equations of motion, thereby deriving result equivalents to Eq.~\eqref{conventionalSET-20269211800}. If the backreaction terms are restored, the results of Eq.~\eqref{LLSE-20269212309} are recovered. Therefore, there is ambiguity in the definition of the conventional GW stress-energy tensor, whereas the GW stress-energy tensor defined in our work yields consistent results across all the aforementioned methods.
\par
Second, in contrast to the conventional GW stress-energy tensor, which is generally not conserved, the GW stress-energy tensor defined in our work is conserved. From identity~\eqref{identity-20266291909} and Eq.~\eqref{GWseT1-20267221052}, the ordinary divergence of the GW stress-energy tensors is readily obtained as:
\begin{eqnarray}
  \partial_{\mu}t^{\mu\nu}=\frac{c^4}{8\pi G}\partial_{\mu}\left\langle -E^{(2)\mu}\bar{A}^{(0)\nu} \right\rangle,\qquad \partial_{\mu}\tau^{\mu\nu}=0.
\end{eqnarray}
The right-hand side of the first equation is generally nonzero. This means an inconsistency between the energy (or momentum) and the energy (or momentum) flux of GWs for the conventional GW stress-energy tensor:
\begin{eqnarray}
  \partial_{0}\int_{\Sigma} t^{0\nu} d^{3}x \ne -\oint_{\partial\Sigma} t^{i \nu} d s_{I},
\end{eqnarray}
where $\Sigma$ denotes the integration volume, $\partial\Sigma$ denotes the boundary of $\Sigma$, and $d s_{I}$ is its outdiffeomorphism vector surface element. In the study of GWs, this is disastrous, because it means that, when investigating the energy (or momentum) change of a system, one cannot a priori use the energy (or momentum) flux carried away by GWs at the boundary as a substitute. In previous study~\cite{Saffer:2017ywl}, the energy (or momentum) flux has been used directly in place of the change in energy (or momentum), even in the absence of an energy conservation relation.
\par
Third, different definitions ($S\left[ A_{\mu},\cdots \right]$ or $S\left[ A^{\mu},\cdots \right]$) of the action yield different results for the conventional GW stress-energy tensor. For the conventional GW stress-energy tensor, the results depend on the definition of the action. Combining Eq.~\eqref{GWseT0-20267221052}, we can obtain:
\begin{eqnarray}
  t^{\mu\nu}-t'^{\mu\nu}=\frac{c^4}{8\pi G}\left\langle S^{(2)\mu\nu}-G^{(2)\mu\nu} \right\rangle - \frac{c^4}{8\pi G}\left\langle S'^{(2)\mu\nu}-G^{(2)\mu\nu} \right\rangle=-\frac{c^4}{4\pi G}\left\langle E^{(2)(\mu}\bar{A}^{(0)\nu)} \right\rangle.
\end{eqnarray}
For the GW stress-energy tensor defined in this work, by Eqs.~\eqref{GWseT0-20267221052},~\eqref{GWseT1-20267221052}, and~\eqref{GWseT2-20267221052}, it is easy to see that different definitions of the action yield the same GW stress-energy tensor. Since Ref.~\cite{Saffer:2017ywl} adopts definition $S\left[ A^{\mu},\cdots \right]$ by default, their energy (or momentum) flux is identical to that obtained from the GW stress-energy tensor defined in our work. However, if definition $S\left[ A_{\mu},\cdots \right]$ is adopted instead, marked differences arise,
\begin{eqnarray}
  \text{For }S\left[ A_{\mu},\cdots \right]:&\quad& \tau^{i\nu}=\frac{c^4}{8\pi G}\left\langle S^{(2)i\nu}-G^{(2)i\nu}-E^{(2)i}\bar{A}\delta_{0}^{\;\nu} \right\rangle \ne t^{i\nu}=\frac{c^4}{8\pi G}\left\langle S^{(2)i\nu}-G^{(2)i\nu} \right\rangle,
  \\
  \text{For }S\left[ A^{\mu},\cdots \right]:&\quad& \tau^{i\nu}=\frac{c^4}{8\pi G}\left\langle S'^{(2)i\nu}-G^{(2)i\nu} \right\rangle =t'^{i\nu}.
\end{eqnarray}
Here, we have substituted the explicit values of the background vector field~\eqref{BackgroundValue-20269221651}. The correction to $\tau^{i\nu}$ vanishes in the contravariant representation because $\bar A^{(0)i}=0$. Since the indices of $\tau^{\mu\nu}$ are not symmetric, the flux of GWs is represented by $\tau^{i\mu}$ rather than $\tau^{\mu i}$.
\par
In summary, for pure metric theories and scalar-tensor theories, the conventional GW stress-energy tensor and the GW stress-energy tensor defined in this work yield the same results. However, for vector-tensor theories with a nonzero background vector, the conventional GW stress-energy tensor is symmetric in its indices but is not conserved, whereas the GW stress-energy tensor defined in this work is conserved but is not symmetric in its indices. A closer analysis shows the following. First, for the conventional GW stress-energy tensor, different methods yield inconsistent results, whereas the GW stress-energy tensor defined in this work yields consistent results under these methods. Second, the nonconservation of the conventional GW stress-energy tensor makes the flux of GWs at the boundary inconsistent with the system's energy and momentum. Since the GW stress-energy tensor defined in this work is conserved, this problem does not arise. Furthermore, for different definitions of the action ($S\left[ A_{\mu},\cdots \right]$ and $S\left[ A^{\mu},\cdots \right]$), the conventional GW stress-energy tensor yields completely different results, whereas the GW stress-energy tensor defined in this work yields consistent results. Finally, for the definition $S\left[ A_{\mu},\cdots \right]$, the conventional GW stress-energy tensor and the GW stress-energy tensor defined in this work yield different fluxes, while for the definition $S\left[ A^{\mu},\cdots \right]$ they yield the same flux.

\par

\section{Einstein-\ae ther theory}\label{Einstein-aether}
Einstein-\ae ther theory provides a concrete test of the conservation framework developed above: a preferred timelike vector modifies both the propagating modes and the complete source that determines their radiation. We first identify the tensor, vector, and scalar sectors and their constraints, then evaluate the completed GW stress-energy tensor and the leading wave zone multipoles. An effective point-particle description subsequently connects these radiative quantities to the losses of a sensitivity-dependent quasicircular binary. Choosing $u_\mu$ as the fundamental vector also permits a comparison with the contravariant-vector and Noether charge calculations of Refs.~\cite{Foster:2007gr,Zhang:2019iim,Lin:2018ken}.

\subsection{Field equations}
Given that Einstein-\ae ther theory is a vector-tensor theory, in strong-field regimes it cannot be assumed a priori that there is no coupling between matter and the vector field. Rather, it is widely recognized in the physics community that, since the extra vector field also constitutes part of the gravitational interaction, its contribution to the binding energy of compact objects may be nonnegligible~\cite{Eardley:1975fgi,Will:1989sk}. Therefore, the action for Einstein-\ae ther theory can be written as~\cite{Jacobson:2000xp,Jacobson:2004ts,Eling:2003rd,Foster:2005dk}
\begin{eqnarray}
  S\left[ g_{\mu\nu},u_{\mu},\lambda,\psi_{m} \right]&=&\frac{c^3}{16\pi G_{0}} \int d^{4}x\sqrt{-g}\left[ R-K^{\alpha\beta\gamma\rho}\nabla_{\alpha}u_{\gamma}\nabla_{\beta}u_{\rho}+\lambda\left( u^{\alpha}u_{\alpha}+1 \right) \right] \nonumber\\
  &&+\frac{1}{c}\int d^4x \sqrt{-g}\mathcal{L}_{m}[g_{\mu\nu},u_{\mu},\psi_{m}], \label{aether-action-20269231730}
\end{eqnarray}
where $G_{0}$ is the bare gravitational constant, $u_{\mu}$ is the \ae ther field, $K^{\alpha\beta\gamma\rho}=c_{1}g^{\alpha\beta}g^{\gamma\rho}+c_{2}g^{\alpha\gamma}g^{\beta\rho}+c_{3}g^{\alpha\rho}g^{\beta\gamma}-c_{4}u^{\alpha}u^{\beta}g^{\gamma\rho}$, with dimensionless couplings $c_1,\ldots,c_4$. The scalar Lagrange multiplier $\lambda$ enforces $u^\alpha u_\alpha=-1$. It transforms as a scalar in the diffeomorphism identity but is not an additional propagating scalar field.
\par
Varying the action~\eqref{aether-action-20269231730} with respect to the multiplier $\lambda$, the vector $u_{\mu}$, and the metric $g_{\mu\nu}$ gives, respectively, the normalization constraint, the vector equation, and the metric equation:
\begin{eqnarray}
  && E_{\phi} = 0,\quad E_{\phi}\equiv \frac{1}{2}\left( u^{\alpha}u_{\alpha}+1 \right), \label{aether-Eq1-20269232238}
  \\
  && E^{\mu} = \frac{8\pi G_{0}}{c^4}J_{m}^{\mu},\quad E^{\mu}\equiv \nabla_{\nu}\left( K^{\nu\alpha\mu\rho}\nabla_{\alpha}u_{\rho} \right) +c_{4}u^{\alpha}(\nabla_{\alpha}u_{\rho})\nabla^{\mu}u^{\rho} +\lambda u^{\mu}, \label{aether-Eq2-20269232238}
  \\
  && G^{\mu\nu} -S^{\mu\nu} = \frac{8\pi G_{0}}{c^4}T_{m}^{\mu\nu},\quad S^{\mu\nu}\equiv  S'^{\mu\nu} -2 E^{(\mu}u^{\nu)} . \label{aether-Eq3-20269232238}
\end{eqnarray}
Here, $T_{m}^{\mu\nu}\equiv \tfrac{2}{\sqrt{-g}}\tfrac{\delta (\sqrt{-g}\mathcal{L}_{m})}{\delta (g_{\mu\nu})}$ and $J_{m}^{\mu}\equiv -\tfrac{1}{\sqrt{-g}}\tfrac{\delta (\sqrt{-g}\mathcal{L}_{m})}{\delta (u_{\mu})}$. The explicit form of $S'^{\mu\nu}$ can be found in, for example, Eq.~(2.14) of Ref.~\cite{Zhang:2019iim} or Eq.~(2.10) of Ref.~\cite{Lin:2018ken}. When comparing expressions, note that those references use $K^{(\mu\nu)}\equiv K^{\mu\nu}+K^{\nu\mu}$, whereas our symmetrization convention includes a factor of $1/2$.
\par
In this section, we study the gravitational radiation of Einstein-\ae ther theory in a Minkowski background. We expand the fields about the Minkowski background:
\begin{eqnarray}
  && g_{\mu\nu}=\eta_{\mu\nu}  + \epsilon h_{\mu\nu},\label{Aether-perturbation1-2026761850}\\
  && u_{\mu}=\bar{u}^{(0)}_{\mu}  + \epsilon \delta u_{\mu},\label{Aether-perturbation2-2026761850}\\
  && \lambda =\bar{\lambda}^{(0)} + \epsilon \Phi.\label{Aether-perturbation3-2026761850}
\end{eqnarray}
Here, $h_{\mu\nu}$, $\delta u_{\mu}$, and $\Phi$ denote all components deviating from the Minkowski background, including the high-frequency perturbations and low-frequency oscillatory components of the background in the Isaacson picture, as well as the static deviation component in the near-source region.
\par
Substituting the perturbation variables~\eqref{Aether-perturbation1-2026761850}-\eqref{Aether-perturbation3-2026761850} into the equations of motion~\eqref{aether-Eq1-20269232238}-\eqref{aether-Eq3-20269232238} and collecting terms at order $\epsilon^0$ yields the background constraints $\bar{u}^{(0)}_{\mu}\bar{u}^{(0)\mu}=-1$ and $\bar{\lambda}^{(0)}=0$. We work in the rest frame of the background \ae ther, choosing
\begin{eqnarray}
  \bar{u}^{(0)}_{\mu}=(-1,0,0,0),\qquad \bar{\lambda}^{(0)}=0.\label{aether-background-20269241554}
\end{eqnarray}
Matter is treated as a perturbative source, $T_{m}^{\mu\nu}=\epsilon T_{m}^{(1)\mu\nu}$ and $J_{m}^{\mu}=\epsilon J_{m}^{(1)\mu}$, with vanishing background values. We therefore use $T_{m}^{\mu\nu}$ and $J_{m}^{\mu}$ for their perturbative parts without ambiguity. Moving the nonlinear gravitational terms to the right-hand side organizes the field equations into linear operators acting on the perturbations and effective sources:
\begin{eqnarray}
  && E^{(\ell)}_{\phi} = 0, \label{aether-PerEq1-20269232238}
  \\
  && E^{(1)\mu} = \frac{8\pi G_{0}}{c^4}\left( J_{m}^{\mu}-\frac{c^4}{8\pi G_{0}}\sum_{\ell \ge 2}E^{(\ell)\mu} \right), \label{aether-PerEq2-20269232238}
  \\
  && G^{(1)\mu\nu} -S^{(1)\mu\nu} = \frac{8\pi G_{0}}{c^4}\left( T_{m}^{\mu\nu} +\frac{c^4}{8\pi G_{0}}\sum_{\ell \ge 2}\left( S^{(\ell)\mu\nu}-G^{(\ell)\mu\nu} \right) \right). \label{aether-PerEq3-20269232238}
\end{eqnarray}
Retaining the quadratic gravitational contributions, we define the effective sources as
\begin{eqnarray}
  J^{\mu}&\equiv& J_{m}^{\mu}-\frac{c^4}{8\pi G_{0}}E^{(2)\mu} +\mathcal{O}(\epsilon^3),\\
  T^{\mu\nu}&\equiv& T_{m}^{\mu\nu} +\frac{c^4}{8\pi G_{0}}\left( S^{(2)\mu\nu}-G^{(2)\mu\nu} \right)+\mathcal{O}(\epsilon^3).
\end{eqnarray}
The metric source $T^{\mu\nu}$ alone is not the conserved quantity relating radiation to source evolution. Specializing the identities~\eqref{matter1-20261022142} and~\eqref{matter2-20261022142} to the background~\eqref{aether-background-20269241554} gives the complete source balance law:
\begin{eqnarray}
  \partial_{\mu}\left( T^{\mu\nu}-J^{\mu}\delta^{\nu}_{\; 0} \right)=0.\label{aether-matter-20269261825}
\end{eqnarray}
This completed source is the starting point for the multipole relations and energy balance below. The quadratic terms supply the leading wave backreaction and contribute to the source moments. Since our analysis is restricted to the quadrupole order of GWs, it suffices to retain the nonlinear source terms up to second order in the perturbations.
\par
Having perturbed the variables~\eqref{Aether-perturbation1-2026761850}-\eqref{Aether-perturbation3-2026761850} and obtained the background constraint~\eqref{aether-background-20269241554}, we now proceed to study the linear perturbations in $\epsilon$ around the background constraint. Since the background spacetime preserves the spatial rotational symmetry, the perturbation variables can be decomposed into scalar, vector, and tensor sectors according to the irreducible representations of the three-dimensional rotation group. At linear order, these sectors evolve independently and can therefore be analyzed separately. According to the scalar-vector-tensor decomposition~\cite{Flanagan:2005yc}, the metric perturbation $h_{\mu\nu}$, the \ae ther-field perturbation $\delta u_{\mu}$, and the sources $T_{\mu\nu}$ and $J_{\mu}$ are uniquely decomposed into the following pieces~\cite{Lai:2024fza,Lai:2026vhe}:
\begin{eqnarray}
  && h_{00}=-2\phi_h, \quad h_{0i}=\lambda_i +\partial_i \varphi_{h},\quad h_{ij}=h^{\mathrm{TT}}_{ij}+\frac{1}{3}\delta_{ij}\theta_{h}+2\partial_{(i}\varepsilon_{j)}+\left( \partial_i\partial_j - \tfrac{1}{3}\delta_{ij}\partial^2\right)\alpha_{h}, \label{SVT1-20266201032}\\
  && \delta u_{0}=\phi_{a},\quad \delta u_{i}=\Sigma_{i}+\partial_{i}\varphi_{a}, \label{SVT2-20266201032}\\
  && T_{00}=\Phi_{m},\quad T_{0i}=\Gamma_{i}+\partial_{i}\Psi_{m}, \quad T_{ij}=T^{\mathrm{TT}}_{ij}+\frac{1}{3}\delta_{ij}\Theta_{m}+2\partial_{(i}\Pi_{j)}+\left( \partial_i\partial_j - \tfrac{1}{3}\delta_{ij}\partial^2\right)\Omega_{m},\label{SVT3-20266201032}\\
  && J_{0}=\tilde{\Phi}_{m},\quad J_{i}=\tilde{\Gamma}_{i}+\partial_{i}\tilde{\Psi}_{m}. \label{SVT4-20266201032}
\end{eqnarray}
Here, $\partial^2=\delta^{ij}\partial_{i}\partial_{j}$, $\delta^{ij}h^{\mathrm{TT}}_{ij}=\delta^{ij}T^{\mathrm{TT}}_{ij}=0$, $\partial^{i}h^{\mathrm{TT}}_{ij}=\partial^{i}T^{\mathrm{TT}}_{ij}=0$, and $\partial^{i}\lambda_i=\partial^{i}\varepsilon_{i} =\partial^{i}\Sigma_{i} =\partial^{i}\Gamma_{i} =\partial^{i}\Pi_{i} =\partial^{i}\tilde{\Gamma}_{i} =0$. 
\par
Since the theory possesses diffeomorphism invariance, it is invariant under infinitesimal coordinate transformations at linear order in $\epsilon$. This implies that we can construct a set of gauge invariants to describe the dynamics of the theory at linear order. Following the infinitesimal coordinate transformations $x^{\mu}\rightarrow x^{\mu}+\xi^{\mu}$, we may construct the following set of gauge invariants~\cite{Lai:2024fza}:
\begin{eqnarray}
  && h^{\mathrm{TT}}_{ij},\; T^{\mathrm{TT}}_{ij},\; \Sigma_{i},\; \Gamma_{i},\; \Pi_{j},\; \tilde{\Gamma}_{i},\; \Phi,\; \Phi_{m},\; \Psi_{m},\; \Theta_{m},\; \Omega_{m},\; \tilde{\Phi}_{m},\; \tilde{\Psi}_{m}, \label{GaugeInvariants1-20266201032}\\
  && \Xi_{i}\equiv \lambda_i-\tfrac{1}{c}\dot{\varepsilon}_{i},\quad \Phi_{h}\equiv \phi_h+\tfrac{1}{c}\dot{\varphi}_{h}-\tfrac{1}{2 c^2}\ddot{\alpha}_{h}, \quad \Theta_{h}\equiv \theta_{h}-\partial^{2}\alpha_{h},\label{GaugeInvariants2-20266201032}\\
  && \Phi_{a}\equiv \phi_{a}-\tfrac{1}{c}\dot{\varphi}_{h}+\tfrac{1}{2 c^2}\ddot{\alpha}_{h},\quad \Psi_{a}\equiv \varphi_{a}-\varphi_{h}+\tfrac{1}{2c}\dot{\alpha}_{h},\label{GaugeInvariants3-20266201032}
\end{eqnarray}
where $\dot{K}=\partial_{t}K=c\partial_{0}K$ and $\partial^{i}\Xi_{i}=0$. The source perturbations are gauge invariant at linear order because their background values vanish~\cite{Lin:2018ken}. Using these gauge invariants, we can express the second-order and first-order perturbative actions, as well as the linear perturbation equations, in a form that does not explicitly contain gauge degrees of freedom.
\par
Substituting the scalar-vector-tensor decomposition~\eqref{SVT1-20266201032}-\eqref{SVT4-20266201032} into the field equations~\eqref{aether-PerEq1-20269232238}-\eqref{aether-PerEq3-20269232238} and retaining terms up to linear order, we obtain the linear perturbation equations. Furthermore, by employing gauge-invariant variables~\eqref{GaugeInvariants1-20266201032}-\eqref{GaugeInvariants3-20266201032} and manipulating the linear perturbation equations, we derive the wave equations and the constraint equations:
\begin{eqnarray}
  && \square_{T}h^{\mathrm{TT}}_{ij} = -\frac{16\pi G_{0}}{c^4} T^{\mathrm{TT}}_{ij},\label{aether-waveEq1-20269251043}
    \\
  && \square_{V}\Sigma_{i} = -\frac{16\pi G_{0}}{c^4(2 c_{1}-c_{+}c_{-})}\left( c_{+}\Gamma_{i} -(1-c_{+})\tilde{\Gamma}_{i} \right),\label{aether-waveEq2-20269251043}
    \\
  && \square_{S}\Theta_{h} = -\frac{16\pi G_{0}}{c^4(2-c_{14})} \left( 3\left(\Phi_{m}+\tilde{\Phi}_{m}\right) -\frac{c_{14}}{c_{123}}\left( (1-c_{+})\Theta_{m}+(2+3c_{2}+c_{+})\partial^2\Omega_{m} \right) \right),\label{aether-waveEq3-20269251043}
    \\
  && \partial^{2}\left( \Xi_{i}+\frac{c_{+}}{1-c_{+}}\Sigma_{i} \right) = -\frac{16\pi G_{0}}{c^4(1-c_{+})} \Gamma_{i},\label{aether-waveEq4-20269251043}
    \\
  && \Phi_{a} = -\Phi_{h}, \quad \partial^2\left( c_{14}(1-c_{+})\Phi_{h}-\tfrac{2 c_{+}-c_{14}}{6}\Theta_{h} \right) = \frac{8\pi G_{0}}{c^4}\left( c_{+}(\Phi_{m}+\tilde{\Phi}_{m})-c_{14}\partial^2\Omega_{m} \right),\label{aether-waveEq5-20269251043}
    \\
  && \partial^2\left( c_{14}(1-c_{+})\partial_{0}\Psi_{a} +\tfrac{2-c_{14}}{6}\Theta_{h} \right) = -\frac{8\pi G_{0}}{c^4}\left( \Phi_{m}+\tilde{\Phi}_{m} -c_{14}\partial^2\Omega_{m} \right),\label{aether-waveEq6-20269251043}
    \\
  && c_{14}(2+2 c_{2}+c_{123})\Phi +\frac{2(c_{23}+c_{2}c_{3})-c_{2}c_{14}+c_{3}c_{123}}{3}\partial^2\Theta_{h} \nonumber\\
   &&\qquad = -\frac{8\pi G_{0}}{c^4}\left( c_{2}c_{14}\Theta_{m} +(2(c_{23}+c_{2}c_{3})+c_{3}c_{123})\Phi_{m} +(2 c_{1234}+c_{134}(c_{+}+3 c_{2}))\tilde{\Phi}_{m} \right).\label{aether-waveEq7-20269251043}
\end{eqnarray}
Here, $c_{+}=c_1+c_3$, $c_{-}=c_{1}-c_{3}$, $c_{l_{1}\cdots l_{n}}=c_{l_{1}}+\cdots +c_{l_{n}}$, $\square_{B}=\tfrac{c^2}{c_{B}^2}\partial^{0}\partial_{0}+\partial^2$ with $B=\{T,V,S\}$. The quantities
$c_{T}^{2}=\tfrac{c^2}{1-c_{+}}$, $c_{V}^{2}=\tfrac{2 c_{1}-c_{+}c_{-}}{2 c_{14} (1-c_{+})}c^2$, and $c_{S}^{2}=\tfrac{c_{123}(2-c_{14})c^2}{c_{14}(1-c_{+})(2+3c_{2}+c_{+})}$ are the squared propagation speeds of the tensor, vector, and scalar sectors, respectively. Equations~\eqref{aether-waveEq1-20269251043}-\eqref{aether-waveEq3-20269251043} describe the five propagating degrees of freedom: two in $h^{\mathrm{TT}}_{ij}$, two in $\Sigma_i$, and one in $\Theta_h$. Equations~\eqref{aether-waveEq4-20269251043}-\eqref{aether-waveEq7-20269251043} are the constraint equations that determine the remaining variables. It is readily seen that, outside the source, the radiative parts of $\Xi_i$ and $\{\Phi_h,\Phi_a,\Psi_a,\Phi\}$ are determined by $\Sigma_i$ and $\Theta_h$ and do not constitute independent radiation channels.
\par
The interpretation of these solutions as physical radiation requires positive kinetic coefficients and hyperbolic propagation with $c_T^2,c_V^2,c_S^2>0$. The coefficients and denominators used below must remain finite. GW170817/GRB170817A constrains the tensor speed at the $10^{-15}$ level under the usual assumptions about the emission delay~\cite{LIGOScientific:2017zic}, requiring $c_+=c_{13}$ to be correspondingly close to zero~\cite{Oost:2018tcv,Baker:2017hug,Dong:2026xab}. This event alone does not impose the same bound on the vector and scalar speeds. Those sectors must be assessed using the complementary requirements from stability, preferred-frame parameterized post-Newtonian (PPN) parameters, binary-pulsar timing, polarization measurements, and gravitational-Cherenkov bounds, see Refs.~\cite{LIGOScientific:2018dkp,Yagi:2013ava}.

\subsection{GW stress-energy tensor in Einstein-\ae ther theory}
In Sec.~\ref{GWSET-20269251143}, we presented four different methods (the perturbed field equation method, the Landau-Lifshitz method, the perturbed action method, and the Noether current method) for defining the GW stress-energy tensor in a general metric theory. In this subsection, we apply them to Einstein-\ae ther theory to obtain an abstract form of the GW stress-energy tensor, retaining terms up to quadrupole order.
\par
By substituting the action~\eqref{aether-action-20269231730}, the equations of motion~\eqref{aether-PerEq1-20269232238}-\eqref{aether-PerEq3-20269232238}, the perturbation~\eqref{Aether-perturbation1-2026761850}-\eqref{Aether-perturbation3-2026761850} and~\eqref{SVT1-20266201032}-\eqref{SVT4-20266201032}, and the gauge invariants~\eqref{GaugeInvariants1-20266201032}-\eqref{GaugeInvariants3-20266201032} into the abstract energy-momentum tensors derived from the four methods (see Eq.~\eqref{GWSET-N-2026771443}, \eqref{GWSET-A-2026771443}, \eqref{perField-20267232110}, and \eqref{LLSE-20269212309} in Sec.~\ref{GWSET-20269251143}), and then using the constraints~\eqref{aether-waveEq4-20269251043}-\eqref{aether-waveEq7-20269251043} and discarding boundary terms, one obtains a consistent GW stress-energy tensor $\tau^{\mu\nu}=\tau_{T}^{\mu\nu}+\tau_{V}^{\mu\nu}+\tau_{S}^{\mu\nu}$ with
\begin{eqnarray}
  \tau_{T}^{\mu\nu} &=& \frac{c^4}{32\pi G_{0}} \left\langle \partial^{\mu}h^{\mathrm{TT}}_{ij} \partial^{\nu}h_{\mathrm{TT}}^{ij} +c_{+}\delta^{\mu}_{\; 0}\partial^{\nu}h^{\mathrm{TT}}_{ij} \partial_{0}h_{\mathrm{TT}}^{ij} \right\rangle,
  \\
  \tau_{V}^{\mu\nu} &=& \frac{c^4}{16\pi G_{0} (1-c_{+})} \left\langle (2c_{1}-c_{+}c_{-})\partial^{\mu}\Sigma_{i}\partial^{\nu}\Sigma^{i} +(c_{+}^{2}-2c_{4}(1-c_{+}))\delta^{\mu}_{\; 0}\partial^{\nu}\Sigma_{i}\partial_{0}\Sigma^{i} \right\rangle,
  \\
  \tau_{S}^{\mu\nu} &=& \frac{c^4}{144\pi G_{0} c_{14}} \left\langle (2-c_{14})\partial^{\mu}\Theta_{h}\partial^{\nu}\Theta_{h} +\tfrac{2(c_{23}-c_{4})+c_{14}(c_{+}^{2}+3c_{2}c_{+}-4 c_{2})}{c_{123}}\delta^{\mu}_{\; 0}\partial^{\nu}\Theta_{h}\partial_{0}\Theta_{h} \right\rangle.
\end{eqnarray}
The GW stress-energy tensors defined by the above four methods generally differ before short-wavelength averaging (i.e., before discarding boundary terms). However, after short-wavelength averaging, they yield the same tensor. It is generally nonsymmetric, as expected in the presence of the preferred timelike background. Moreover, by definition (see Sec.~\ref{GWSET-20269251143}), the conservation law holds only for the first index, $\partial_{\mu}\tau^{\mu\nu}=0$.
\par
We now proceed to derive the GW solutions via the multipole expansion in the asymptotic region at infinity. The retarded Green's function for Eq.~\eqref{aether-waveEq1-20269251043} gives the tensor solution
\begin{eqnarray}
  h^{\mathrm{TT}}_{ij}(x)=\frac{4 G_{0}}{c^4}\int d^{3}\vec{x}' \frac{1}{|\vec{x}-\vec{x}'|}T^{\mathrm{TT}}_{ij}\left(x^{0}-\tfrac{c}{c_{T}}|\vec{x}-\vec{x}'|,\vec{x}'\right).
\end{eqnarray}
Typically, the characteristic velocity of the source $v_{s}$ is sufficiently small compared with the speed of GWs, i.e., $v_{s}/c_{T}\ll 1$. Moreover, the detector is located at a distance far from the source, i.e., $R\gg d$, where $R$ denotes the distance between the source and the detector and $d$ denotes the characteristic length scale of the source. We may therefore perform the far-field and nonrelativistic expansion of the solution. Retaining terms up to $1/R$ order, the solution can be written as:
\begin{eqnarray}
  h^{\mathrm{TT}}_{ij}(x)\approx \frac{4 G_{0}}{c^4R}\hat{\Lambda}_{ij,kl}\sum_{n=0}^{\infty}\frac{c^n}{n!c_{T}^n}\partial_{0}^{n}\left[ \int d^{3}\vec{x}'T^{kl}\left(x^{0}-\tfrac{c}{c_{T}}R,\vec{x}'\right)\left( \vec{x}'\cdot\vec{N} \right)^n \right]. \label{expansionT-20269252252}
\end{eqnarray}
Here, $\vec{N}$ denotes the unit spatial vector directed from the source to the detector, $\hat{\Lambda}_{ij,kl}\big(\vec{N}\big)\equiv \hat{P}_{ik}\hat{P}_{jl}-\tfrac{1}{2}\hat{P}_{ij}\hat{P}_{kl}$ is the transverse-traceless operator, and $\hat{P}_{ij}=\hat{P}_{ij}\big(\vec{N}\big)$ is the transverse operator. With $x'^i\sim d$, $N^i\sim1$, and $\partial_0=c^{-1}\partial_t\sim \omega/c$, where $\omega$ is the characteristic source frequency and $\omega d\sim v_s$, the different multipole moments of $h^{\mathrm{TT}}_{ij}$ in Eq.~\eqref{expansionT-20269252252} satisfy the following order-of-magnitude relations:
\begin{eqnarray}
  \frac{h^{\mathrm{TT}}_{ij}\big |_{n=n_{0}+1}}{h^{\mathrm{TT}}_{ij}\big |_{n=n_{0}}} \sim\mathcal{O}\left( v_{s}/c_{T} \right).
\end{eqnarray}
The expansion index $n$ counts powers of the retardation correction. Let us now examine the leading order of $h^{\mathrm{TT}}_{ij}$ which is the quadrupole order, namely the term with $n=0$,
\begin{eqnarray}
  h^{\mathrm{TT}}_{ij}|_{n=0}=\frac{2 G_{0}}{c^4R}\hat{\Lambda}_{ij,kl}\left[ \ddot{M}^{kl}+\frac{2}{c}\dot{\mathcal{M}}^{(k,l)} \right].\label{aether-solution0-20269262125}
\end{eqnarray}
The source moments are $M^{l_{1}\cdots l_{n}}\equiv \frac{1}{c^2}\int d^3\vec{x}'\left(T^{00}-J^{0}\right)x'^{l_{1}}\cdots x'^{l_{n}}$ and $\mathcal{M}^{k,l_{1}\cdots l_{n}}\equiv \int d^3\vec{x}'J^{k}x'^{l_{1}}\cdots x'^{l_{n}}$. The appearance of both moments reflects the vector contribution to the complete conserved source and is essential when relating the spatial stress to its radiative moments.
\par
Applying the same retarded construction to Eqs.~\eqref{aether-waveEq2-20269251043} and~\eqref{aether-waveEq3-20269251043} gives the vector and scalar solutions. Their expansions require $v_s/c_V\ll1$ and $v_s/c_S\ll1$, respectively:
\begin{eqnarray}
  \Sigma_{i}(x)&\approx& -\frac{4 G_{0}}{c^4R(2c_{1}-c_{+}c_{-})}\hat{P}_{ij}\sum^{\infty}_{n=0}\frac{c^n}{n!c_{V}^n}\partial_{0}^{n}\left[ \int d^{3}\vec{x}'\left( c_{+}(T^{0j}-J^{j})+J^{j} \right)\left( \vec{x}'\cdot\vec{N} \right)^n \right],
  \\
  \Theta_{h}(x)&\approx& \frac{4 G_{0}}{c^4R(2-c_{14})}\sum^{\infty}_{n=0}\frac{c^n}{n!c_{S}^n}\partial_{0}^{n}\Bigg[ \int d^{3}\vec{x}'\Big( 3\left(T^{00}-J^{0}\right) +\frac{3c_{14}}{2}T^{i}_{\; i} \nonumber\\
  && -\frac{3 c_{14}(2+c_{+}+3c_{2})}{2c_{123}}N^{i}N^{j}T_{ij} \Big)\left( \vec{x}'\cdot\vec{N} \right)^n \Bigg].
\end{eqnarray}
Retaining both the vector and scalar wave solutions up to the quadrupole order, including the scalar trace contribution, gives
\begin{eqnarray}
  \Sigma_{i}|_{n=0}&=&\frac{4 G_{0}(c_{+}-1)}{c^4 R(2c_{1}-c_{+}c_{-})}\hat{P}_{ij}\mathcal{M}^{j}+\text{const.},\label{aether-solution1-20269262125}
  \\
  \Sigma_{i}|_{n=1}&=&-\frac{2 G_{0}}{c^3 R c_{V}(2c_{1}-c_{+}c_{-})}\hat{P}_{ij}N_{k}\left( c_{+}\ddot{M}^{jk}+\tfrac{2c_{+}}{c}\dot{\mathcal{M}}^{(j,k)}+\tfrac{2(1-c_{+})}{c}\dot{\mathcal{M}}^{j,k} \right),\label{aether-solution2-20269262125}
  \\
  \Theta_{h}|_{n=0}&=& -\frac{ G_{0} c_{14}}{c^4 R c_{123}(2-c_{14})}\left( -3c_{123}\delta_{ij}+3(2+3c_{2}+c_{+})N_{i}N_{j} \right)\left( \ddot{M}^{ij}+\tfrac{2}{c}\dot{\mathcal{M}}^{(i,j)} \right)+\text{const.}, \label{aether-solution3-20269262125}
  \\
  \Theta_{h}|_{n=1}&=& \frac{ 12 G_{0} }{c^2 R c_{S}(2-c_{14})}N_{i}\dot{M}^{i}, \label{aether-solution4-20269262125}
  \\
  \Theta_{h}|_{n=2}&=& \frac{ 6 G_{0} }{c^2 R c_{S}^2(2-c_{14})}N_{i}N_{j}\ddot{M}^{i j}. \label{aether-solution5-20269262125}
\end{eqnarray}
The complete source conservation law~\eqref{aether-matter-20269261825} makes $\int d^3x'\,T^{0j}$ and $\int d^3x'\,(T^{00}-J^0)$ constant at the retained order when boundary terms vanish. Their time-independent contributions are denoted by the constants in Eqs.~\eqref{aether-solution1-20269262125} and~\eqref{aether-solution3-20269262125} and do not carry radiative flux. In Eqs.~\eqref{aether-solution4-20269262125} and~\eqref{aether-solution5-20269262125}, we neglect higher-order terms beyond the quadrupole order. 
\par
The GW stress-energy tensor now connects the outgoing solutions to energy transport. We define the radiative energy loss as $\dot{E}$. The vacuum wave-zone conservation law determines this flux through $\tau^{i0}$. Relating it to the source energy requires the complete conserved source, including matter and nonlinear near-zone fields. Thus,
\begin{eqnarray}
  \dot{E} &=& \partial_{t}\int_{\Sigma}d^{3}\vec{x}\tau^{00}=-c\int_{\Sigma}d^{3}\vec{x}\partial_{k}\tau^{k0}=-c\oint_{\partial\Sigma}dS N_{k}\tau^{k0}\nonumber
  \\
  &=& -\frac{c^5}{32\pi G_{0}}\oint_{\partial\Sigma}dS N_{k} \left\langle \partial^{k}h^{\mathrm{TT}}_{ij} \partial^{0}h_{\mathrm{TT}}^{ij} +\tfrac{2(2c_{1}-c_{+}c_{-})}{1-c_{+}}\partial^{k}\Sigma_{i}\partial^{0}\Sigma^{i} +\tfrac{2(2-c_{14})}{9c_{14}}\partial^{k}\Theta_{h}\partial^{0}\Theta_{h} \right\rangle\nonumber
  \\
  &\approx& -\frac{c^4}{32\pi G_{0}}\oint_{\partial\Sigma}dS \left\langle \tfrac{1}{c_{T}}\dot{h}^{\mathrm{TT}}_{ij} \dot{h}_{\mathrm{TT}}^{ij} +\tfrac{2(2c_{1}-c_{+}c_{-})}{(1-c_{+})c_{V}}\dot{\Sigma}_{i}\dot{\Sigma}^{i} +\tfrac{2(2-c_{14})}{9c_{14}c_{S}}\dot{\Theta}_{h}\dot{\Theta}_{h} \right\rangle.\label{aether-dE-20269262113}
\end{eqnarray}
Here, we have applied the relation $N_{i}\partial^{i}\left(\tfrac{1}{R}K(t-\tfrac{R}{c_{B}})\right)=-\tfrac{1}{c_{B}}\partial_{t}\left(\tfrac{1}{R}K(t-\tfrac{R}{c_{B}})\right) +\mathcal{O}(1/R^2)$. From the first line of the above equation~\eqref{aether-dE-20269262113}, it is readily seen that the GW stress-energy tensor defined in this work ensures consistency between the radiative energy loss and the energy flux at the boundary. The specific form of $h^{\mathrm{TT}}_{ij}$, $\Sigma_{i}$, and $\Theta_{h}$ is given in Eqs.~\eqref{aether-solution0-20269262125} and~\eqref{aether-solution1-20269262125}-\eqref{aether-solution5-20269262125}.
\par
As can be seen from the solutions~\eqref{aether-solution0-20269262125} and~\eqref{aether-solution1-20269262125}-\eqref{aether-solution5-20269262125}, the leading-order contributions of the vector and scalar gravitational radiation are dipole radiation, in sharp contrast to tensor gravitational radiation, whose leading order is quadrupole radiation. This implies that, for suitable source models, vector and scalar gravitational radiation in Einstein-\ae ther theory can be much stronger than tensor gravitational radiation. However, in GW detection, one must also consider whether such source models can actually exist in the Universe.

\subsection{The binary system in Einstein-\ae ther theory}\label{BS-Eae}
In the study of gravitational radiation in Einstein-\ae ther theory, we focus here on compact binary systems. Because the radiation is considered in the far zone, sufficiently far from the source, the compact bodies can be effectively modeled as point particles. In vector-tensor theories, however, the vector field is also an integral component of the gravitational interaction. Since compact bodies reside in the strong-field regime, the vector field's contribution to the particle mass cannot be neglected a priori~~\cite{Eardley:1975fgi,Will:1989sk,Zhang:2017sym,Higashino:2022izi,Quartin:2023tpl,Foster:2007gr,Lin:2018ken,Zhang:2019iim,Taherasghari:2022wfs,Taherasghari:2023rwn,Taherasghari:2025mlf}. In this subsection, we investigate the gravitational radiation emitted by binary systems within this framework based on the action $S\left[ u_{\mu},\cdots \right]$ and choose the gauge $\varphi_h=\alpha_h=0$.
\par
For a body $I$, the mass function $m_I[\gamma_I]$ describes the dependence of its binding energy on motion relative to the local \ae ther. The effective matter action is~\cite{Foster:2007gr,Lin:2018ken,Zhang:2019iim,Taherasghari:2023rwn,Taherasghari:2025mlf}
\begin{eqnarray}
  S_{m}=-\sum_{I}\int d\tau_{I}\; m_{I}[\gamma_{I}]c^2,\label{aether-matter-20269281011}
\end{eqnarray}
where $\tau_{I}$ is the proper time of the body $I$, $\gamma_{I}=-u_{\mu}v^{\mu}_{I}/c$ measures its motion relative to the \ae ther, and $v^{\mu}_{I}$ denotes the four-velocity of the corresponding body. In what follows, we shall use $\gamma$ as a shorthand notation for $\gamma_{I}$. In the nonrelativistic limit relative to the background \ae ther, $\gamma\to1$, so the mass function can be expanded as~\cite{Foster:2007gr,Lin:2018ken,Zhang:2019iim,Taherasghari:2025mlf}
\begin{eqnarray}
  m_{I}[\gamma]=m_{I}\left( 1 +s_{I}(\gamma-1) +\frac{1}{2}(s_{I}^2-s_{I}+s'_{I})(\gamma-1)^2 +\mathcal{O}\left((\gamma -1)^3\right) \right).
\end{eqnarray}
Here, $m_{I}\equiv m_{I}[\gamma]|_{\gamma =1}$ is the reference mass, and the dimensionless sensitivity parameters are
\begin{eqnarray}
  s_{I}\equiv \frac{d \mathrm{ln} m_{I}[\gamma]}{d \mathrm{ln} \gamma}\Big{|}_{\gamma=1},\quad s'_{I}\equiv \frac{d^2 \mathrm{ln} m_{I}[\gamma]}{d (\mathrm{ln} \gamma)^2}\Big{|}_{\gamma=1}.
\end{eqnarray}
The sensitivities $s_I$ and $s'_I$ carry information about each body's internal gravitational structure. For a given body, they can be determined by considering asymptotic properties of perturbations of static solutions.
\par
In order to obtain the matter stress-energy tensor and the matter current, we vary the matter action~\eqref{aether-matter-20269281011} with respect to the metric $g_{\mu\nu}$ and the \ae ther field $u_{\mu}$, respectively,
\begin{eqnarray}
  \frac{\delta S_{m}}{\delta g_{\mu\nu}} &=& \sum_{I}\int d^4 x\; \frac{1}{2} \left( m_{I}[\gamma]-\gamma m_{I,\gamma}[\gamma] \right)\frac{v_{I}^{\mu}v_{I}^{\nu}}{v_{I}^{0}}\delta^{(3)}(\vec{x}-\vec{x}_{I}),
  \\
  \frac{\delta S_{m}}{\delta u_{\mu}} &=& \sum_{I}\int d^4 x\; m_{I,\gamma}[\gamma] c \frac{v_{I}^{\mu}}{v_{I}^{0}}\delta^{(3)}(\vec{x}-\vec{x}_{I}).
\end{eqnarray}
Using the definitions~\eqref{Eq1-20266162330} and~\eqref{Eq2-20266162330}, we therefore obtain the matter stress-energy tensor and vector current:
\begin{eqnarray}
  T_{m}^{\mu\nu}&=&\frac{c}{\sqrt{-g}}\sum_{I}\left( m_{I}[\gamma]-\gamma m_{I,\gamma}[\gamma] \right)\frac{v_{I}^{\mu}v_{I}^{\nu}}{v_{I}^{0}}\delta^{(3)}(\vec{x}-\vec{x}_{I}),\label{aether-matter1-20269281526}
  \\
  J_{m}^{\mu}&=&-\frac{c^2}{\sqrt{-g}}\sum_{I}m_{I,\gamma}[\gamma]\frac{v_{I}^{\mu}}{v_{I}^{0}}\delta^{(3)}(\vec{x}-\vec{x}_{I}).\label{aether-matter2-20269281526}
\end{eqnarray}
Rewriting the same action in terms of $u^\mu$ changes the metric variation because the contravariant, rather than covariant, vector components are then held fixed. The resulting matter stress-energy tensor has a different expression, see Eq.~(2.10) of Ref.~\cite{Taherasghari:2025mlf}. This change of field variables leaves the physical theory unchanged.
\par
Varying the matter action~\eqref{aether-matter-20269281011} with respect to the trajectory gives the geodesic equation of motion, $\delta S_m/\delta x_I^\mu=0$:
\begin{eqnarray}
  m^{*}_{I}[\gamma]v_{I}^{\alpha}\nabla_{\alpha}v_{I}^{\mu} = m_{I,\gamma\gamma}[\gamma]\left( \gamma v_{I}^{\mu} -u^{\mu}c \right)v_{I}^{\alpha}\partial_{\alpha}\gamma +2 m_{I,\gamma}[\gamma]cg^{\mu\nu}v_{I}^{\alpha} \partial_{[\nu}u_{\alpha]},\label{aether-geodesic-20269281125}
\end{eqnarray}
where $m^{*}_{I}[\gamma]=m_{I}[\gamma]-\gamma m_{I,\gamma}[\gamma]$. The geodesic equation is independent of the choice between action definitions $S\left[ u_{\mu},\cdots \right]$ and $S\left[ u^{\mu},\cdots \right]$, and is consistent with the matter identities~\eqref{identityMatter-20266291909} and~\eqref{identityMatter1-20266291909}. To obtain the geodesic equation applicable at the quadrupole order for gravitational radiation, we need to retain the geodesic equation up to order $v_{I}^2$. This implies that it suffices to retain it to Newtonian order (higher-order geodesic equation is given in Eq. (2.21) of Ref.~\cite{Taherasghari:2025mlf}). At Newtonian order, the background spacetime is static, Eq.~\eqref{aether-geodesic-20269281125} reduces to
\begin{eqnarray}
  \frac{d^2x_{I}^{i}}{dt^2} +c^2\partial^{i}\phi_{h} - \frac{m_{I,\gamma}c^2}{m^{*}_{I}}\partial^{i}\phi_{a}=0.\label{aether-geodesic1-20269281211}
\end{eqnarray}
Here, we have used that the leading-order static solutions of the gravitational field variables are $\mathcal{O}(v_{I}^2)$. Since only $\phi_{h}$ and $\phi_{a}$ appear in Eq.~\eqref{aether-geodesic1-20269281211}, we give below only the explicit static solution for $\phi_{h}$ and $\phi_{a}$. In the Newtonian limit, the wave equation of $\Theta_{h}$~\eqref{aether-waveEq3-20269251043} reduces to Poisson's equation. Inserting the latter into Eq.~\eqref{aether-waveEq5-20269251043} yields the static solution for $\phi_{h}$ and $\phi_{a}$ after considering the gauge $\varphi_{h}=\alpha_{h}=0$:
\begin{eqnarray}
  -\phi_{a}=\phi_{h} &=& -\frac{2 G_{0}}{c^4(2-c_{14})}\int d^{3}\vec{x}'\frac{1}{|\vec{x}-\vec{x}'|}(T^{00}-J^{0}+T^{i}_{\;\; i})\nonumber\\
  &\approx& -\frac{2 G_{0}}{c^2(2-c_{14})}\sum_{I}\frac{m_{I}}{|\vec{x}-\vec{x}_{I}|},\label{aether-static-20269281539}
\end{eqnarray}
In the second line, we have inserted Eqs.~\eqref{aether-matter1-20269281526} and~\eqref{aether-matter2-20269281526} and retained only Newtonian terms. Substituting the above static solution~\eqref{aether-static-20269281539} into the geodesic equation~\eqref{aether-geodesic1-20269281211}, with body $1$ and body $2$ acting as mutual gravitational sources in the binary system, we obtain the equations of motion for the two bodies:
\begin{eqnarray}
  \frac{d^2 x_{1}^{i}}{d t^2}=-\frac{\tilde{G}\tilde{m}_{2}}{r^3}r^{i}, \quad \frac{d^2 x_{2}^{i}}{d t^2}=\frac{\tilde{G}\tilde{m}_{1}}{r^3}r^{i}.
\end{eqnarray}
Here, $r^i=x_{1}^{i}-x_{2}^{i}$ is the relative separation, $\tilde{G}=\tfrac{2 G_{0}}{(2-c_{14})(1-s_{1})(1-s_{2})}$ is the effective gravitational coupling between the bodies, and $\tilde{m}_{I}=(1-s_{I})m_{I}$ is the effective inertial mass. For a quasicircular orbit in the center-of-mass frame, $\tilde{m}_{1}x_{1}^{i}+\tilde{m}_{2}x_{2}^{i}=0$, the positions and relative speed satisfy
\begin{eqnarray}
  x_{1}^{i}=\frac{\tilde{\mu}}{\tilde{m}_{1}}r^{i},\quad x_{2}^{i}=-\frac{\tilde{\mu}}{\tilde{m}_{2}}r^{i},\quad v^2=\frac{\tilde{G}\tilde{m}}{r}, \label{aether-relation-20269281619}
\end{eqnarray}
where $\tilde{\mu}=\tfrac{\tilde{m}_{1}\tilde{m}_{2}}{\tilde{m}_{1}+\tilde{m}_{2}}$ is the reduced effective mass, $\tilde{m}=\tilde{m}_{1}+\tilde{m}_{2}$ is the total effective inertial mass, and $v^{i}=\dot{r}^{i}$ is the relative velocity. In Eq.~\eqref{aether-relation-20269281619}, the constant-radius condition for quasicircular orbits is used, namely $\dot{r}=0$. The binary barycenter is assumed to be at rest in the asymptotic \ae ther frame, and $1-s_I>0$ is required for positive inertial masses. For preferred-frame effects from a moving barycenter, see Ref.~\cite{Zhang:2019iim}.
\par
\begin{figure}[tbp]
  \centering
  \includegraphics[width=0.78\textwidth]{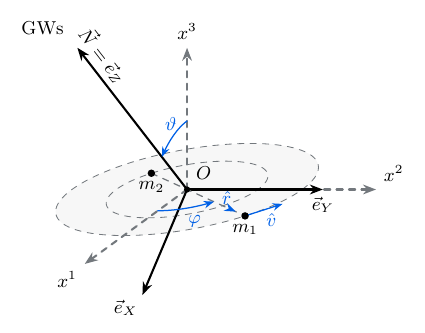}
  \caption{Schematic of the quasicircular binary and its outgoing GWs. The thick dashed arrows and thick solid arrows denote the coordinate system $\{x^1,x^2,x^3\}$ and the orthogonal basis $\{\vec e_X,\vec e_Y,\vec e_Z\}$, respectively. The bodies orbit their common center of mass $O$ in the $x^1$-$x^2$ plane. Although GWs are emitted in all directions, this schematic shows only those propagating along the $\vec N$ direction.}
  \label{fig:binary-orbit-geometry}
\end{figure}

Substituting the matter stress-energy tensor~\eqref{aether-matter1-20269281526}, the matter current~\eqref{aether-matter2-20269281526}, and the relations~\eqref{aether-relation-20269281619} into the GW solutions~\eqref{aether-solution0-20269262125} and~\eqref{aether-solution1-20269262125}-\eqref{aether-solution5-20269262125}, we can obtain the GW solutions for a quasicircular orbit in the center-of-mass frame of
the binary,
\begin{eqnarray}
  h^{\mathrm{TT}}_{ij}|_{n=0}&=&\frac{4 G_{0}\tilde{G}\tilde{m}\tilde{\mu}}{c^4R r}\hat{\Lambda}_{ij,kl}\left( \hat{v}^{k}\hat{v}^{l}-\hat{r}^{k}\hat{r}^{l} \right),\label{aether-solution1-2026929912}
  \\
  \Sigma_{i}|_{n=0}&=&\frac{4 G_{0}\tilde{\mu}}{c^3 R}\left( \tfrac{s_{1}}{1-s_{1}}-\tfrac{s_{2}}{1-s_{2}} \right)\frac{1-c_{+}}{2 c_{1}-c_{+}c_{-}}\hat{P}_{ij}v^{j},\label{aether-solution2-2026929912}
  \\
  \Sigma_{i}|_{n=1}&=&\frac{4 G_{0}\tilde{G}\tilde{\mu}}{c^3 R r c_{V}(2c_{1}-c_{+}c_{-})}\left( \tfrac{s_{1}-c_{+}}{1-s_{1}}\tilde{m}_{2}+\tfrac{s_{2}-c_{+}}{1-s_{2}}\tilde{m}_{1} \right)\hat{P}_{ij}N_{k}\left( \hat{v}^{j}\hat{v}^{k}-\hat{r}^{j}\hat{r}^{k} \right),\label{aether-solution3-2026929912}
  \\
  \Theta_{h}|_{n=0}&=& -\frac{6 G_{0}\tilde{G}\tilde{m}\tilde{\mu}}{c^4 R r}\frac{c_{14}(2+3c_{2}+c_{+})}{c_{123}(2-c_{14})}N_{i}N_{j}\left( \hat{v}^{i}\hat{v}^{j}-\hat{r}^{i}\hat{r}^{j} \right), \label{aether-solution4-2026929912}
  \\
  \Theta_{h}|_{n=1}&=& \frac{ 6 \tilde{G} \tilde{\mu} }{c^2 R c_{S}}(s_{1}-s_{2})N_{i}v^{i}, \label{aether-solution5-2026929912}
  \\
  \Theta_{h}|_{n=2}&=& \frac{ 6\tilde{G}^2\tilde{\mu} }{c^2 R r c_{S}^2}\left( (1-s_{1})\tilde{m}_{1} +(1-s_{2})\tilde{m}_{2} \right) N_{i}N_{j}\left( \hat{v}^{i}\hat{v}^{j}-\hat{r}^{i}\hat{r}^{j} \right),\label{aether-solution6-2026929912}
\end{eqnarray}
where $\hat{r}^{i}=r^{i}/|r^{i}|$ and $\hat{v}^{i}=v^{i}/|v^{i}|$ are unit vectors. We assume that the orbit of the binary system lies in the $x^{1}-x^{2}$ plane of the coordinate system $\{x^{1},x^{2},x^{3}\}$. The system is then axially symmetric about the $x^{3}$-axis. Hence, without loss of generality, we take $N^{i}$ to lie in the $x^{1}-x^{3}$ plane and to make an angle $\mathcal{\vartheta}$ with the $x^{3}$-axis. Accordingly, we introduce an orthogonal basis $\{\vec{e}_{X},\vec{e}_{Y},\vec{e}_{Z}\}$ with $\vec e_Z=\vec N$ and $\vec e_Y$ aligned with the original $x^2$ axis (see Fig.~\ref{fig:binary-orbit-geometry}). Therefore, in the coordinate system $\{x^{1},x^{2},x^{3}\}$, the relevant key vectors are given by
\begin{eqnarray}
  && \vec{e}_{X}=(\mathrm{cos}\vartheta, 0, -\mathrm{sin}\vartheta),\quad \vec{e}_{Y}=(0,1,0),\quad \vec{e}_{Z}=\vec{N}=(\mathrm{sin}\vartheta, 0, \mathrm{cos}\vartheta),\label{aether-coordinate1-2026929912}\\
  && \hat{r}=(\mathrm{cos}\varphi, \mathrm{sin}\varphi, 0),\quad \hat{v}=(-\mathrm{sin}\varphi, \mathrm{cos}\varphi, 0),\label{aether-coordinate2-2026929912}
\end{eqnarray}
where $\varphi$ is the orbital phase measured from the $x^1$ axis. 
Substituting these fields and angular relations into Eq.~\eqref{aether-dE-20269262113}, integrating over the sphere, and averaging over an orbit period gives
\begin{eqnarray}
  \dot{E} &=& -\frac{2G_{N}^3m_{1}^2m_{2}^2}{3 r^4}\frac{(s_{1}-s_{2})^2}{(1-s_{1})^2(1-s_{2})^2 c_{14}}\left( \frac{1}{c_{S}^3} +\frac{2-c_{14}}{c_{V}^3} \right) \nonumber\\
  && -\frac{8 G_{N}^3m_{1}^2m_{2}^2}{5 r^4}\frac{\tilde{G}\tilde{m}}{r} \Bigg[\frac{2(2-c_{14})}{c^4c_{T}} +\frac{2-c_{14}}{c_{14}(1-c_{+})^2c_{V}^5}\left( \frac{s_{2}-c_{+}}{1-s_{2}}\frac{\tilde{m}_{1}}{\tilde{m}} +\frac{s_{1}-c_{+}}{1-s_{1}}\frac{\tilde{m}_{2}}{\tilde{m}} \right)^2\nonumber\\
  && +\frac{4 }{3c_{14}c_{S}^5} \left( \frac{\tilde{m}_{1}}{(1-s_{2})\tilde{m}} +\frac{\tilde{m}_{2}}{(1-s_{1})\tilde{m}} - \frac{2-c_{14}}{2(1-c_{+})} \right)^2 \Bigg]. \label{aether-dE-20269301824}
\end{eqnarray}
Here, $G_{N}=2G_{0}/(2-c_{14})$ is the Newtonian gravitational constant, and the orbital radius is treated as constant during the average. The first term is the combined scalar and vector dipole loss, while the remaining terms are the tensor, vector, and scalar quadrupole losses. To exhibit the order-of-magnitude relations, define $\Delta s=s_1-s_2$ and collect the remaining dimensionless factors into finite weights $C_D$ and $C_Q$:
\begin{eqnarray}
  \dot{E}_{\text{dipole}} &\sim& C_D\frac{G_{N}^3m_{1}^2m_{2}^2}{ r^4 c^3} (\Delta s)^2,\label{aether-dE1-20269301824}
  \\
  \dot{E}_{\text{quadrupole}} &\sim& C_Q\frac{G_{N}^3m_{1}^2m_{2}^2}{ r^4 c^3}\frac{v^2}{c^2},\label{aether-dE2-20269301824}
\end{eqnarray}
where we have used $v^2=\tilde{G}\tilde{m}/r$. The weights contain the coupling parameters, $(1-s_I)$ factors, and speed ratios $c_B/c$ with $B={T,V,S}$. The dipole enters at relative $-1$PN order with respect to the quadrupole, giving a flux ratio that scales as $(C_D/C_Q)(\Delta s)^2(c/v)^2$. The contributions become comparable when $|\Delta s|\sim(v/c)\sqrt{C_Q/C_D}$, while equal sensitivities eliminate the leading dipole. These scalings assume $v/c_B\ll1$ in every radiating sector. A quantitative estimate must use compact-body sensitivities and couplings compatible with the existing constraints~\cite{Yagi:2013ava}.
\par
Equation~\eqref{aether-dE-20269301824} connects the conserved radiative flux to binary evolution. For an adiabatic quasicircular inspiral, a consistently matched balance law gives $\dot\Omega=\dot E/(dE_{\mathrm{orb}}/d\Omega)$, where $E_{\mathrm{orb}}$ is the binary orbital energy and $\Omega=\dot\varphi$ is the orbital angular frequency. The sensitivity-dependent dipole term enters at relative $-1$PN order and can therefore be constrained through orbital decay in binaries with unequal sensitivities, particularly neutron-star-white-dwarf systems~\cite{Yagi:2013ava,Gupta:2021vdj}. The same additional losses modify the accumulated inspiral phase, providing a complementary observable for LISA, TianQin, and Taiji~\cite{LISA:2022kgy,TianQin:2015yph,Hu:2017mde}. 
\par
Gravitational radiation in Einstein-aether theory has been studied in Refs.~\cite{Foster:2007gr,Zhang:2019iim,Taherasghari:2025mlf}. However, the results reported there are not mutually consistent. We will discuss the similarities and differences between our results and those obtained in these works. The angular projectors $\{N_{i},\hat{P}_{ij},\hat{\Lambda}_{ij,kl}\}$ in Eq.~\eqref{aether-dE-20269262113} can be integrated using the identities in Chap.~3 of Ref.~\cite{Maggiore2007}. For the point-particle sources introduced in Eqs.~\eqref{aether-matter1-20269281526} and~\eqref{aether-matter2-20269281526}, the integrated result has the form of Eq.~(3.42) of Ref.~\cite{Zhang:2019iim}, with a difference in the vector part of $\mathcal A_2$. A comparison requires consistent speed, sensitivity, and mass conventions. We use dimensionless speeds $\hat c_B=c_B/c$ and identify our sensitivity with $s_I=-\sigma_I$ in that reference. Our $m_I$ is its mass function evaluated at $\gamma_I=1$, while $\tilde m_I=(1-s_I)m_I$ is the effective inertial mass. Our coefficient is
\begin{eqnarray}
  \mathcal A_2=\frac{2c_{+}}{(2c_1-c_{+}c_-)\hat c_V^3}-\frac{2(Z-1)}{(2-c_{14})\hat c_S^3}, \qquad
  Z=\frac{(\alpha_1-2\alpha_2)(1-c_{+})}{3(2c_{+}-c_{14})},
\end{eqnarray}
where $\alpha_1$ and $\alpha_2$ are the preferred-frame PPN parameters, see Ref.~\cite{Zhang:2019iim}. The original printed coefficient in Ref.~\cite{Zhang:2019iim} has the opposite sign in its vector term. Comparing it with that reference's variational definition of the matter current gives $T_\mu=-\sum_A\tilde m_A\tilde\delta_A A_A^2v_\mu^A$, rather than the positive sign printed in Eq.~(2.14). Its conserved source in Eq.~(3.5) correspondingly reads $\tau^{\mu\nu}=T^{\mu\nu}+T^\mu\delta^\nu_{\; 0}$. These signs restore agreement with the vector contribution obtained here. Furthermore, the general expression for our GW energy-loss rate~\eqref{aether-dE-20269262113} agrees with Eq.~(71) of Ref.~\cite{Foster:2007gr}. In its explicit multipole form, the tensor and vector contributions, together with the scalar dipole contribution, agree with Foster's results, whereas the scalar quadrupole contribution differs somewhat in its coefficients. The above differences are minor and may be attributable to typographical errors. Next, we focus on the controversial differences among studies reported in Ref.~\cite{Taherasghari:2025mlf}. In particular, our results differ significantly from those of Ref.~\cite{Taherasghari:2025mlf}. This difference is directly visible by comparing the binary-system result~\eqref{aether-dE-20269301824} above with Eqs. (4.7) and (5.3) of Ref.~\cite{Taherasghari:2025mlf}.
\par
Taherasghari and Will investigated the orbital energy loss of compact binaries in Einstein-\AE ther theory using near-zone radiation reaction in Paper 1~\cite{Taherasghari:2023rwn} and Paper 2~\cite{Taherasghari:2025mlf}. Paper 1 constructs PN solutions for the near-zone fields, while Paper 2 combines these solutions with a modified geodesic equation to derive the binary accelerations and the orbital energy-loss rate. Regarding the comparison with published studies, they state: ``We find significant disagreements with published results, based on calculating the energy flux in the far zone using a `Noether current' construction.''~\cite{Taherasghari:2025mlf}. This discrepancy motivates a consistency check of whether the field redefinition and near-zone truncation in Paper 1 retain the contributions of all physical modes through the required radiation-reaction order. For a mode with propagation speed $v_{\bullet}$, the wave operator is $\partial^2-v_{\bullet}^{-2}\partial_t^2$, with the following relations:
\begin{eqnarray}
  \text{Near zone:} && \partial^{2} \;\sim\; \frac{1}{\ell_{s}^2},\quad \frac{1}{v_{\bullet}^2}\partial_{t}^2 \;\sim\; \frac{\omega_{GW}^{2}}{v_{\bullet}^2},
  \\
  \text{Wave zone:} && \partial^{2} \;\sim\; \frac{1}{\ell_{GW}^2},\quad \frac{1}{v_{\bullet}^2}\partial_{t}^2 \;\sim\; \frac{\omega_{GW}^{2}}{v_{\bullet}^2}.
\end{eqnarray}
Here, $\ell_{s}$, $\ell_{GW}$, and $\omega_{GW}$ are the characteristic spatial scale of the near-zone gravitational potential, the characteristic wavelength of GWs, and the characteristic frequency of the radiation, respectively. Therefore, the ratio between the characteristic magnitudes of the temporal and spatial derivative terms is as follows:
\begin{eqnarray}
    \text{Near zone:} && \left| \frac{\omega_{GW}\ell_{s}}{v_{\bullet}} \right| \ll 1,
    \\
    \text{Wave zone:} && \left| \frac{\omega_{GW}\ell_{GW}}{v_{\bullet}} \right| \sim 1.
\end{eqnarray}
Note that the GW speeds in Refs.~\cite{Taherasghari:2023rwn,Taherasghari:2025mlf} are denoted by $v_{\bullet}$: $v_T$ and $v_L$ for the vector and scalar modes, respectively, and the speed of light for the tensor mode when $c_{13}=0$. The near-zone expansion requires $|\omega_{\mathrm{GW}}\ell_s/v_{\bullet}|\ll1$ for every relevant mode, which implies that radiation reaction arises from higher-order terms in the retarded solution. The suppression of temporal derivatives does not by itself imply a loss of radiative information. The question is whether these terms are retained to the required order after the field redefinition and truncation.
\par
For $c_{13}=0$, the leading dipole term in Eq.~\eqref{aether-dE-20269301824} agrees with Foster's result quoted in Eq.~(5.3) of Paper 2. Using the dimensionless speeds $v_L=c_S/c$ and $v_T=c_V/c$, with $v_T^2=c_1/c_{14}$, its coefficient after factoring out the common mass, sensitivity, and orbital dependence is $2/(c_{14}v_L^3)+2(2-c_{14})/(c_1v_T)$, whereas the corresponding coefficient in Eq.~(4.7) of Paper 2 is $1+3(2-c_{14})/(c_1v_T)$. Thus, the discrepancy is already explicit at leading dipole order: their result lacks the $v_L^{-3}$ dependence, and the numerical factor multiplying $(2-c_{14})/(c_1v_T)$ is $3$ rather than $2$. In our calculation, the scalar dipole amplitude~\eqref{aether-solution5-2026929912}, combined with the outgoing-flux expression~\eqref{aether-dE-20269262113}, gives the $v_L^{-3}$ dependence. These two differences motivate examining how the near-zone calculation retains the scalar and vector propagation information.
\par
Specifically, Eqs.~(4.13) of Paper 1 form a finite-order near-zone PN system obtained through the field transformation (4.7). The new variables cannot be directly identified with the physical radiative eigenmodes in the wave zone. Two possible sources of the speed-dependent discrepancy require examination:
\begin{itemize}
\item I. The scalars among the new variables satisfy wave equations with unit propagation speed, whereas Appendix A of Paper 1, as well as the present work and Refs.~\cite{Foster:2007gr,Zhang:2019iim}, gives the physical scalar GW speed as $v_L$. It is therefore necessary to check how the $v_L$ dependence is recovered when the original fields are reconstructed at the PN order relevant to radiation reaction.
\item II. The spatial potential retains a longitudinal scalar component in addition to the transverse vector component, while its truncated wave equation involves $v_T$. If these components are not consistently separated after field reconstruction, a longitudinal contribution could be assigned the vector propagation speed and alter the coefficient of the $v_T$-dependent term.
\end{itemize}
Treating the finite-order near-zone equations as complete wave-zone equations exceeds their established domain of validity. The relevant test is whether the original variables reconstructed from Eqs.~(4.7) and~(4.13) satisfy the original field equations to the stated PN accuracy. Resolving these issues requires an explicit matching of the near-zone solution to the wave-zone scalar and vector modes at the PN order relevant to radiation reaction. Such a calculation is needed to determine whether the field redefinition, truncation, and mode decomposition account for the two dipole-coefficient differences. A comparison at higher orders must also include PN corrections to the dipole moment and other nonlinear contributions.

\section{Conclusion}\label{Conclusion-sec}
In this work, we investigated the construction of a conserved GW stress-energy tensor in local, diffeomorphism-invariant metric theories containing vector and scalar fields, motivated by the need for a complete energy-momentum balance when additional gravitational degrees of freedom contribute to radiation. Guided by Isaacson's interpretation of GW backreaction~\cite{Isaacson:1968hbi,Isaacson:1968zza}, we treated all gravitational fields as a coupled system and used the diffeomorphism identity~\cite{Gralla:2013rwa} to identify the complete conserved source. We examined the perturbed field equation, perturbed action, Landau-Lifshitz, and Noether current methods~\cite{Saffer:2017ywl} at leading Isaacson order in a flat vacuum wave zone with constant background fields. We then applied this framework to Einstein-\ae ther theory, using far-zone multipole expansions and a sensitivity-dependent point-particle description~\cite{Foster:2007gr} to calculate the tensor, vector, and scalar radiative losses of an adiabatic quasicircular compact binary.
\par
At leading Isaacson order in a flat vacuum wave zone with constant background fields, our GW stress-energy tensor agrees with the conventional expression in pure metric and scalar-tensor theories. However, with a nonzero background vector, the conventional symmetric metric source~\cite{Saffer:2017ywl,Heisenberg:2025roe} need not be separately conserved. The required vector-equation contribution completes it into a generally nonsymmetric tensor satisfying $\partial_\mu\tau^{\mu\nu}=0$, with energy flux $c\tau^{i0}$ and a consistent energy-momentum balance. All four constructions then yield equivalent averaged tensors and fluxes on shell, independently of whether $A_\mu$ or $A^\mu$ is fundamental. This independence does not generally hold for the conventional metric source: in the rest frame of a timelike background vector, its flux coincides with the completed flux in the $A^\mu$ formulation but can differ in the $A_\mu$ formulation. In Einstein-\ae ther theory, the general expression for our GW energy-loss rate agrees with Eq.~(71) of Ref.~\cite{Foster:2007gr}. In its explicit multipole form, the tensor and vector contributions, together with the scalar dipole contribution, agree with Foster's results, whereas the scalar quadrupole contribution differs somewhat in its coefficients. We also traced the discrepancy in the vector contribution to $\mathcal{A}_2$ in the original printed expressions of Ref.~\cite{Zhang:2019iim} to sign inconsistencies in the matter current and source conservation law. Correcting these signs restores agreement. In particular, we examined the possible origins of the confusion in Refs.~\cite{Taherasghari:2023rwn,Taherasghari:2025mlf} regarding the discrepancy between the near-zone results of their authors and the wave-zone results previously reported in Refs.~\cite{Foster:2007gr,Zhang:2019iim}, see Sec.~\ref{BS-Eae}.
\par
The nonsymmetry of the conserved GW stress-energy tensor warrants further investigation of its physical interpretation and the freedom to redefine the tensor. A fixed preferred vector can break boost invariance of the perturbation theory while the full theory remains diffeomorphism invariant, so standard Belinfante symmetrization is not automatically applicable~\cite{belinfante1940current,Saffer:2017ywl}. Future work should determine whether an admissible superpotential improvement can yield a symmetric, conserved representative while preserving asymptotic charges and radiative fluxes. It should also clarify the relation to the symmetric metric backreaction source and formulate the appropriate angular-momentum balance, accounting consistently for spin and background-field contributions. A complementary approach in GR~\cite{Koivisto:2026zpz,Koivisto:2026yxt} uses a frame field independent of the metric to define canonical energy-momentum currents. Unlike the dynamical vector considered here, this frame field has an identically vanishing bulk Euler-Lagrange derivative. In a canonical frame~\cite{Koivisto:2026zpz,BeltranJimenez:2019bnx}, the translation currents are conserved for coupled solutions without Killing symmetries. Their superpotentials and, in QCD, spin contributions follow from the variational principle. The relation to our averaged GW stress-energy tensor in dynamical vector theories remains to be established. This will require consistent matching of frame choices, asymptotic charges, and radiative fluxes.
\par
The conserved GW stress-energy tensor developed here provides a foundation for relating the energy carried by additional gravitational degrees of freedom to source evolution and observable waveforms. Combined with consistent source dynamics, wave propagation, and detector responses, this framework can support quantitative tests of modified gravity. Future space-based observatories, including Taiji, TianQin, and LISA~\cite{LISA:2017pwj,Hu:2017mde,TianQin:2015yph}, will open new opportunities for such tests in the millihertz band. Their observations of long-lived inspirals can probe additional radiation channels through accumulated phase shifts, complementing direct searches for extra polarizations~\cite{LISA:2022kgy}. A consistent account of gravitational energy transport can thus help turn precision GW measurements into tests of both the field content and the dynamical laws of gravity.

\section*{Acknowledgments}
AI-assisted tools were used to refine the language and formatting of the LaTeX manuscript. Responsibility for the scientific content and its validation remains with the authors.
\par
We thank Professor Anzhong Wang for discussions and assistance regarding GW energy and gravitational radiation, and Professor Nicolas Yunes for his analysis and suggestions.
This work is supported in part by the National Natural Science Foundation of China (Grants No. 12475056, No. 123B2074, and No. 12247101), Gansu Province's Top Leading Talent Support Plan, the Fundamental Research Funds for the Central Universities (Grant No. lzujbky-2025-jdzx07), the Natural Science Foundation of Gansu Province (No. 22JR5RA389 and No. 25JRRA799), the 111 Project (Grant No. B20063), and the Department of Education of Gansu Province: Outstanding Graduate ``Innovation Star” Project (Grant No. 2026CXZX-044).\par

\appendix
\section{GW stress-energy tensor based on contravariant vector}\label{appendix-GWSET-20269251143}
In Sec.~\ref{GWSET-20269251143} of the main text, we derived the GW stress-energy tensor using the perturbed field equation, the Landau-Lifshitz, the perturbed action, and the Noether current methods, based on the definition $S\left[ A_{\mu},\cdots \right]$. Here, we briefly present the corresponding results obtained using the equivalent definition $S\left[ A^{\mu},\cdots \right]$.

\subsection{The field-theoretical perspective based on contravariant vector}
In the Minkowski spacetime, the quadratic gravitational action, based on the definition $S\left[ A^{\mu},\cdots \right]$, is
\begin{eqnarray}
  S_{g}^{(2)}\left[ h_{\mu\nu},\delta A^{\mu},\delta\phi \right]=\int dtd^{3}\vec{x} \;\tilde{\mathcal{L}}'_{g}\left[ h_{\mu\nu},\delta A^{\mu},\delta\phi \right],\label{appendix-Sg2-2026771258}
\end{eqnarray}
Based on this action, we can define the GW stress-energy tensor via the Noether current method and the perturbed action method, respectively.
\par
\textit{\textbf{The Noether current method.}} For the second-order perturbed action~\eqref{appendix-Sg2-2026771258}, under spacetime translation transformations, one obtains a conserved current $j'^{\mu}_{\;\;\nu}$ which satisfies $\partial_{\mu}j'^{\mu}_{\;\;\nu}=0$~\cite{Noether:1918zz,Olver1993,Iyer:1994ys,Maggiore2007,Saffer:2017ywl}, and its specific form is:
\begin{eqnarray}
  j'^{\mu}_{\;\;\nu}&=& \delta^{\mu}_{\;\;\nu}\tilde{\mathcal{L}}'_{g}-\frac{\partial\tilde{\mathcal{L}}'_{g}}{\partial(\partial_{\mu}h_{\alpha\rho})}\partial_{\nu}h_{\alpha\rho} -\frac{\partial\tilde{\mathcal{L}}'_{g}}{\partial(\partial_{\mu}\delta A^{\alpha})}\partial_{\nu}\delta A^{\alpha} -\frac{\partial\tilde{\mathcal{L}}'_{g}}{\partial(\partial_{\mu}\delta\phi)}\partial_{\nu}\delta\phi +\cdots, \label{appendix-jCSE-2026771442}
\end{eqnarray}
where the ellipsis denotes contributions from derivatives beyond first order. This is known as the canonical stress-energy tensor. Since the energy of GWs cannot be well-defined at a local point in space, the GW stress-energy tensor is typically assumed to be the short-wavelength average of the conserved current,
\begin{eqnarray}
  \tau^{\mu\nu} =\langle j'^{\mu\nu} \rangle. \label{GWSET-N1-2026771443}
\end{eqnarray}
The averaged tensor is generally nonsymmetric and satisfies $\partial_\mu\tau^{\mu\nu}=0$.
\par
\textit{\textbf{The perturbed action method.}} We apply the diffeomorphism variations to the background fields in $S_g^{(2)}$,
\begin{eqnarray}
  \delta_{\xi}\left(S_{g}^{(2)}\right)&=&\frac{1}{c}\int d^4x\sqrt{-\bar{g}}\left[\frac{\delta \left(\tilde{\mathcal{L}}'_{g}\right)}{\sqrt{-\bar{g}}\delta \left(\bar{g}_{\mu\nu}\right)}\delta_{\xi} \left(\bar{g}_{\mu\nu}\right) +\frac{\delta \left(\tilde{\mathcal{L}}'_{g}\right)}{\sqrt{-\bar{g}}\delta \left(\bar{A}^{\mu}\right)}\delta_{\xi} \left(\bar{A}^{\mu}\right) +\frac{\delta \left(\tilde{\mathcal{L}}'_{g}\right)}{\sqrt{-\bar{g}}\delta \left(\bar{\phi}\right)}\delta_{\xi} \left(\bar{\phi}\right) \right]\nonumber\\
  &=& \frac{1}{c}\int d^4x\sqrt{-\bar{g}}\xi_{\nu}\partial_{\mu}\left[ \frac{2}{\sqrt{-\bar{g}}}\frac{\delta \left(\tilde{\mathcal{L}}'_{g}\right)}{\delta \left(\bar{g}_{\mu\nu}\right)} -\frac{\bar{g}^{\nu\alpha}}{\sqrt{-\bar{g}}}\frac{\delta \left(\tilde{\mathcal{L}}'_{g}\right)}{\delta \left(\bar{A}^{\alpha}\right)}\bar{A}^{\mu} \right].\label{appendix-PerAction1-2026772258}
\end{eqnarray}
Diffeomorphism invariance and the arbitrariness of $\xi_\nu$ then imply that the bracket in this equation has vanishing first-index divergence. We therefore define the GW stress-energy tensor as
\begin{eqnarray}
  \tau^{\mu\nu}= \left\langle \frac{2}{\sqrt{-\bar{g}}}\frac{\delta \left(\tilde{\mathcal{L}}'_{g}\right)}{\delta \left(\bar{g}_{\mu\nu}\right)} -\frac{\bar{g}^{\nu\alpha}}{\sqrt{-\bar{g}}}\frac{\delta \left(\tilde{\mathcal{L}}'_{g}\right)}{\delta \left(\bar{A}^{\alpha}\right)}\bar{A}^{\mu} \right\rangle,\label{GWSET-A1-2026771443}
\end{eqnarray}
where the background fields are set to the values in Eq.~\eqref{background31-2026761811} after variation. As in the Noether construction, the resulting GW stress-energy tensor is generally nonsymmetric and conserved on its first index.

\subsection{The geometrical perspective based on contravariant vector}
For the geometrical perspective, an important concept is the Isaacson picture, see Sec.~\ref{IPandQoM-2026761755}. At this point, we can no longer a priori approximate the background spacetime as fixed. Based on the full equations of motion, using the linear wave equations, and retaining terms of backreaction order, we can obtain the linearized and sourced equations of motion for $\{\delta \bar{g}_{\mu\nu},\delta \bar{A}^{\mu},\delta \bar{\phi}\}$,
\begin{eqnarray}
  \bar{G}^{(1)\mu\nu} -\bar{S}'^{(1)\mu\nu} &=& \frac{8\pi G}{c^4}\left[  T_{m}'^{(2)\mu\nu} +\tfrac{c^4}{8\pi G}\big( S'^{(2)\mu\nu}-G^{(2)\mu\nu} \big) \right],\label{appendix-BGEq1-2026781504}\\
  \bar{E}^{(1)\mu} &=& \frac{8\pi G}{c^4} \left[ J_{m}^{(2)\mu} -\tfrac{c^4}{8\pi G}E^{(2)\mu} \right],\label{appendix-BGEq2-2026781504}\\
  \bar{E}^{(1)}_{\phi} &=& \frac{8\pi G}{c^4} \left[ J_{\phi}^{(2)} -\tfrac{c^4}{8\pi G}E^{(2)}_{\phi} \right].\label{appendix-BGEq3-2026781504}
\end{eqnarray}
It is readily seen from these three equations that $G^{(2)\mu\nu}$, $S'^{(2)\mu\nu}$, $E^{(2)\mu}$, and $E^{(2)}_{\phi}$ serve as sources for the slowly varying components of the background.
\par
\textit{\textbf{The perturbed field equation method.}} The vector and scalar equations also have quadratic sources, and the diffeomorphism identity~\eqref{identity1-20266291909} determines how they enter the energy-momentum balance.  Following a procedure similar to that used to derive Eq.~\eqref{perField-20267232110}, we obtain the GW stress-energy tensor
\begin{eqnarray}
  \tau^{\mu\nu}= \frac{c^4}{8\pi G}\left\langle S'^{(2)\mu\nu}-G^{(2)\mu\nu}-E^{(2)\nu}\bar{A}^{(0)\mu} \right\rangle. \label{perField1-20267232110}
\end{eqnarray}
Similar to the GW stress-energy tensors obtained by the previous two methods, these stress-energy tensors are also nonsymmetric, and only their first index satisfies the conservation equation.
\par
\textit{\textbf{The Landau-Lifshitz method.}} The complete Landau-Lifshitz density is conserved before the separation into waves and background. Extracting its radiative part requires retaining the slow background response and its vector equation at the same order as the quadratic wave terms. The constant zeroth-order superpotential contributes no derivatives, and the linear high-frequency term averages to zero. At the leading backreaction order, the remaining identity is
\begin{eqnarray}
  0 &=& \left(\partial_{\alpha}\partial_{\mu}\partial_{\nu}\mathcal{H}^{\alpha\mu\beta\nu}\right)^{(2)} \nonumber\\
  &=& \frac{16\pi G}{c^4}\partial_{\alpha}\left[ (-g) \left( T_{m}'^{\alpha\beta} +J_{m}^{\beta}A^{\alpha} \right) +(-g) \left( t_{LL}^{\alpha\beta} +\tfrac{c^4}{8\pi G}S'^{\alpha\beta} -\tfrac{c^4}{8\pi G}E^{\beta}A^{\alpha} \right) \right]^{(2)}.\label{appendix-HLL-20267142146}
\end{eqnarray}
This rearrangement isolates a completed matter contribution and a completed gravitational contribution with vanishing total ordinary divergence. In the vacuum wave zone, their averaged quadratic gravitational part defines 
\begin{eqnarray}
  \tau^{\mu\nu}=\left\langle \left[(-g) \left( t_{LL}^{\mu\nu} +\tfrac{c^4}{8\pi G}S'^{\mu\nu} -\tfrac{c^4}{8\pi G}E^{\nu}A^{\mu} \right) \right]^{(2)} \right\rangle,
\end{eqnarray}
Combining Eq.~\eqref{ddH-20267232102}, $(-g)t_{LL}^{\mu\nu}$ and $-\tfrac{c^4}{8\pi G}(-g)G^{\mu\nu}$ differ only by a total derivative term, which, under the averaging operation, reduces to a boundary term and can therefore be neglected. Therefore, the GW stress-energy pseudotensor can be rewritten in the tensor form as
\begin{eqnarray}
  \tau^{\mu\nu}=\frac{c^4}{8\pi G}\left\langle S'^{(2)\mu\nu}-G^{(2)\mu\nu}-E^{(2)\nu}\bar{A}^{(0)\mu} \right\rangle. \label{GWSE-202610191208}
\end{eqnarray}
Therefore, the definition of the GW stress-energy tensors obtained by the Landau-Lifshitz method is identical to that given by the perturbed field equation method~\eqref{perField1-20267232110} in the flat vacuum wave zone and at leading order in short-wavelength averaging.

\bibliographystyle{unsrt}
\bibliography{referenceData}

\end{document}